\documentclass[aps,twocolumn,preprintnumbers,amsmath,amssymb,prb,superscriptaddress]{revtex4}
\usepackage[dvips]{epsfig}
\usepackage{subfigure}
\usepackage{graphicx}
\usepackage{longtable}
\usepackage[ansinew]{inputenc}
\usepackage{ulem}
\usepackage{color}

\begin{document}

\title{Signatures of an Atomic Crystal in the Band Structure of a Molecular Thin Film}

\author{Norman Haag}
\affiliation{Department of Physics and Research Center OPTIMAS, University of Kaiserslautern, 67663 Kaiserslautern, Germany}
\author{Daniel L\"uftner}
\affiliation{Institute of Physics, University of Graz, NAWI Graz, Universit\"atsplatz 5, 8010 Graz, Austria}
\author{Florian Haag}
\affiliation{Department of Physics and Research Center OPTIMAS, University of Kaiserslautern, 67663 Kaiserslautern, Germany}
\author{Johannes Seidel}
\affiliation{Department of Physics and Research Center OPTIMAS, University of Kaiserslautern, 67663 Kaiserslautern, Germany}
\author{Leah L. Kelly}
\affiliation{Department of Physics and Research Center OPTIMAS, University of Kaiserslautern, 67663 Kaiserslautern, Germany}
\author{Giovanni Zamborlini}
\affiliation{Experimentelle Physik VI, Technische Universit\"at Dortmund, 44221 Dortmund, Germany}
\affiliation{Peter Gr\"unberg Institut (PGI-6), Forschungszentrum J\"ulich, 52425 J\"ulich, Germany}
\author{Matteo Jugovac}
\affiliation{Peter Gr\"unberg Institut (PGI-6), Forschungszentrum J\"ulich, 52425 J\"ulich, Germany}
\author{Vitaliy Feyer}
\affiliation{Peter Gr\"unberg Institut (PGI-6), Forschungszentrum J\"ulich, 52425 J\"ulich, Germany}
\author{Martin Aeschlimann}
\affiliation{Department of Physics and Research Center OPTIMAS, University of Kaiserslautern, 67663 Kaiserslautern, Germany}
\author{Peter Puschnig}
\affiliation{Institute of Physics, University of Graz, NAWI Graz, Universit\"atsplatz 5, 8010 Graz, Austria}
\author{Mirko Cinchetti}
\affiliation{Experimentelle Physik VI, Technische Universit\"at Dortmund, 44221 Dortmund, Germany}
\author{Benjamin Stadtm\"uller}
\affiliation{Department of Physics and Research Center OPTIMAS, University of Kaiserslautern, 67663 Kaiserslautern, Germany}
\email{bstadtmueller@physik.uni-kl.de}

\date{\today}

\begin{abstract}
Transport phenomena in molecular materials are intrinsically linked to the orbital character and the degree of localization of the valence states. Here, we combine angle-resolved photoemission with photoemission tomography to determine the spatial distribution of all molecular states of the valence band structure of a C$_{60}$ thin film. While the two most frontier valence states exhibit a strong band dispersion, the states at larger binding energies are characterized by distinct emission patterns in energy and momentum space. Our findings demonstrate the formation of an atomic crystal-like band structure in a molecular solid with delocalized $\pi$-like valence states and strongly localized $\sigma$-states at larger binding energies.
\end{abstract}

\maketitle

In the last decades, molecular systems have emerged as highly tuneable materials for optoelectronic, photonic, and spintronic applications with the unique opportunity to actively design and control the optical band gap of light active materials by chemical functionalization \cite{Schwarze.2016, Bizzarri.2017, Xu.2014, Cinchetti.2017}. Despite this intriguing chance for technological applications, the overall efficiency of molecular devices still suffers from the rather low charge carrier mobility and our generally poor understanding of the charge transport mechanisms in molecular solids.  

Both challenges have triggered intensive research focusing on either the chemical synthesis of novel molecular complexes with record charge carrier mobility \cite{Anthony.2006, Wang.2012, Kang.2013,Liu.2015} or the improvement of the models describing charge transport in these materials. So far, it was demonstrated that charge transport in organic materials can range from purely polaron hopping transport to coherent band-like transport depending on the band structure of the material\cite{Ueno.2008,Bredas.2004,Kera.2015, Machida.2010,Bussolotti.2017,Latzke.2019}. While delocalization and pronounced band dispersion of the frontier orbitals, in particular, of the highest occupied molecular orbital (HOMO) and the lowest unoccupied molecular orbital (LUMO), are an important prerequisite for coherent band-like transport, hopping transport usually occurs in molecular materials with valence states localized at the individual molecular sites. Unfortunately, even today, a quantitative understanding of the degree of delocalization of molecular transport levels for different intermolecular interactions is still elusive. This is particularly true for three-dimensional molecular complexes, such as rubrene \cite{ Menard.2004, Sundar.2004} or the C$_{60}$ derivative PCBM \cite{Anthopoulos.2004}, which have demonstrated exceptionally large charge carrier mobility.

In this Letter, we combine angle-resolved photoelectron spectroscopy (ARPES) and photoemission tomography (PT) to determine the spatial localization of all molecular orbitals of the entire valence band structure of the prototypical three-dimensional organic molecule C$_{60}$. In PT, the angle-resolved photoemission yield from a molecular orbital can be interpreted as the Fourier transform of the corresponding real-space molecular wave function \cite{Puschnig.2009}. Despite the simplicity of this model, which roots in the assumption of a plane-wave final state, it has been extremely successful in explaining the ARPES signatures of planar $\pi$-conjugated molecules adsorbed on surfaces \cite{Ziroff.2010, Willenbockel.2013,Feyer.2014,Grimm.2018,Stadtmuller.2012,Weiss.2015,Egger.2019,Zamborlini.2017} and to disentangle the spectroscopic signatures of structural or chemically inequivalent planar molecules in monolayer films on surfaces\cite{Stadtmuller.2012,Stadtmuller.2014,Willenbockel.2015}. Only very recently, the plane wave final state was also applied to predict the ARPES signatures of a monolayer film of non-planar fullerene molecules on a metal surface\cite{Metzger.2020}. 

Here, we focus on the band structure of a thin C$ _{60}$ film on the Ag(111) surface. We find that the two most frontier molecular orbitals reveal a strong band dispersion while the energetically lower lying orbitals appear as distinct emission maxima in momentum space. These differences can be explained by density functional theory (DFT) calculations of free standing C$_{60}$ layer in conjunction with PT simulations. Accordingly, we can assign the strongly dispersing bands to the HOMO and HOMO-1 with pure $\pi$-orbital character which spread over adjacent molecules. In contrast, the sharper emission maxima at larger binding energies are caused by $\sigma$-orbitals that are localized on the individual C$_{60}$ sites. Our findings demonstrate that the band structure of molecular materials can exhibit the same characteristic signatures known from any crystalline inorganic material with delocalized valence and localized (molecular) core level states.

We start with an overview of the molecular band structure of the C$_{60}$ thin film ($5\,$ML) on Ag(111). The C$_{60}$ molecules arrange in a crystalline $(2\sqrt{3}\times 2\sqrt{3})R30^\circ$ structure with two coexisting structural domains which are rotated by $\pm18^\circ$. To suppress any thermally induced rotation of the C$_{60}$ molecules, the sample temperature was kept below $150\,$K throughout the experiment \cite{Dresselhaus.1996}. At this low temperature, molecular motion is suppressed and the C$_{60}$ crystal undergoes a phase transition into a simple cubic phase with four C$_{60}$ molecules per surface unit cell: one hexagon prone and three double-bond prone molecules\cite{David.1992, David.1991}. All photoemission data were acquired with a momentum microscope and synchrotron radiation at the Light Source Elettra, which allows us to record the complete energy and momentum dependent photoemission yield within the photoemission horizon in a fixed experimental geometry \cite{Kromker.2008, Tusche.2013}. More details can be found in the supplementary material (see, also, references \cite{Shi.2012, Tamai.2005} therein). 

\begin{figure}[ht]
	\centering
		\includegraphics[width=8.5cm]{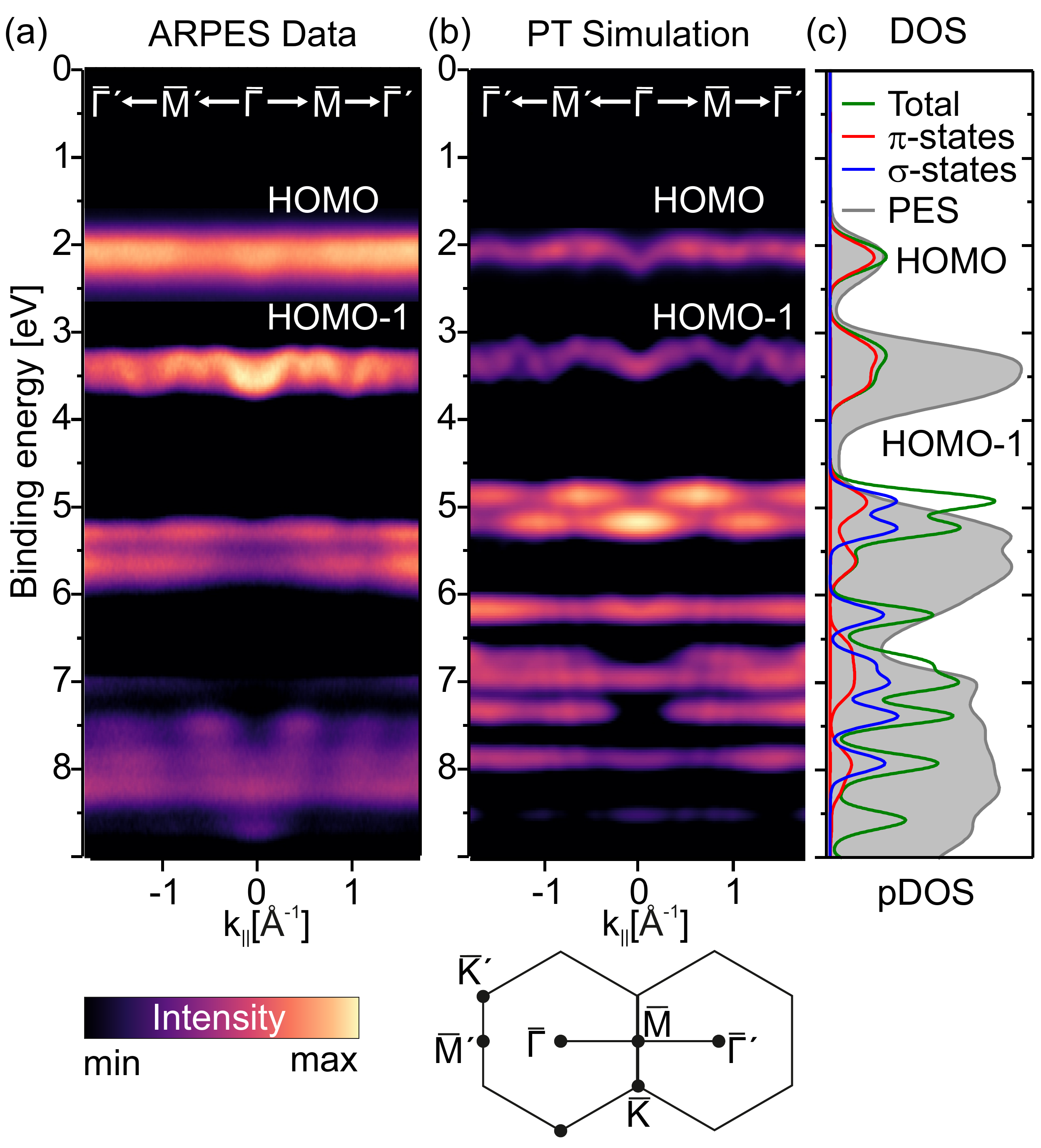}
	\caption{Energy vs. momentum cut through the ARPES data along the $\overline{\Gamma}$~$\overline{\mathrm{M}}$~$\overline{\Gamma}'$-direction of the surface Brillouin zone for a crystalline C$_{60}$ thin film ($5\,$ML, E$_\mathrm{photon}=35\,$eV, p-polarized light). The surface Brillouin zone is shown on the right side of the band structure cut. (b) PT simulation of the same energy vs. momentum cut based on a DFT calculation of a freestanding C$_{60}$ layer. (c) Density of states projected onto the $\pi$- and $\sigma$-states of C$_{60}$. For comparison, the spectral density of the total photoemission yield is included as gray curve.} 
	\label{fig:Fig1}
\end{figure}

Fig.~1a shows the molecular valence band structure of the C$_{60}$ film as an energy vs. momentum cut along the $\overline{\Gamma}$~$\overline{\mathrm{M}}$~$\overline{\Gamma}'$-direction of the surface Brillouin zone of the C$_{60}$ crystal. We find significant differences in the energy and momentum distributions of the molecular features depending on their binding energy. The first two molecular states (E$_{\mathrm{B}}<5\,$eV) show a band dispersion (band width) of up to $700\,$meV. This band width is rather large for molecular materials with predominant van-der-Waals interactions, but still smaller compared to the typical band width of inorganic semiconductors or other inorganic crystalline materials \cite{Hoffmann.1987,Golze.2019}. In contrast, the molecular features at larger binding energies (E$_{\mathrm{B}}>5\,$eV) appear as distinct maxima in energy and momentum space. Therefore, at first glance, the band structure of the C$_{60}$ film reveals all characteristic features of the band structure of an atomic solid with dispersing valence bands and localized (molecular) core level states.

Theoretical insight into the ARPES data can be obtained by the PT simulation based on DFT calculations \cite{Perdew.1996} of a freestanding C$_{60}$ layer with $4$ molecules per unit cell (details can be found in the supplementary material, and references \cite{Kresse1993,Kresse1999,Tkatchenko2009,Bloechl1994} therein). We calculated the 3D Fourier transform of the molecular wave function of each molecular state and extracted the photoemission signal by a spherical cut through the 3D Fourier transform in momentum space \cite{Puschnig.2009}, an approach which has recently been extended to account for 2D dispersion layers \cite{Luftner.2017}. The radius of the spherical momentum space cut is determined by the total momentum k$_\mathrm{final}$ of the electrons in the photoemission final state. For planar molecules on surfaces, k$_\mathrm{final}$ is determined by the kinetic energy of the emitted photoelectrons. In case of our C$_{60}$ film, we additionally need to consider the inner potential V$_0$ which renormalizes the perpendicular component k$_\mathrm{z}$ of the electron momentum in the final state. This is particularly crucial for 3D molecules for which the 3D Fourier transform reveals a strong intensity modulation along the k$_\mathrm{z}$ direction (see also Fig.~6 in the supplementary material). 

For the PT simulation, we used the inner potential of V$_0=13\,$eV, which was determined experimentally by Hasegawa et al. \cite{Hasegawa.1998}. The corresponding energy vs. momentum cut of our PT simulations along the $\overline{\Gamma}$~$\overline{\mathrm{M}}$~$\overline{\Gamma}'$-direction is shown in Fig.~1b. Note that we have aligned the energy of the topmost band with its experimentally observed binding energy position. We find an excellent agreement between the PT simulation and the experimental data (Fig.~1a). The PT simulation reveals two dispersing bands for small binding energies with a comparable band width and energy difference as observed experimentally, and discrete emission features for binding energies larger than E$_\mathrm{B}>5\,$eV. The almost rigid energy shift of the lower-lying states in the simulation with respect to the experiment is typical for the used generalized gradient approximation (GGA) functional \cite{Puschnig.2015} and can be attributed to self-interaction errors.

The overall excellent agreement between experiment and theory allows us to determine the orbital character of all molecular photoemission signatures by projecting the density of states onto the $\pi$- and $\sigma$-orbitals of C$_{60}$. The corresponding projected density of states (pDOS) is shown in Fig.~1c. The pDOS of the two most frontier orbitals is purely dominated by $\pi$-states and can hence be attributed to the HOMO and HOMO-1 bands of C$_{60}$. Importantly, the HOMO band is derived from the five H$_{\mathrm{u}}$ orbitals, the HOMO-1 from the four G$_{\mathrm{g}}$ and five H$_{\mathrm{g}}$ orbitals of the free C$_{60}$ molecule. At larger binding energy, the molecular photoemission signals contain a mixture of $\pi$- and $\sigma$-states. 

The predictive power of our PT simulations becomes even clearer when turning to the constant energy (CE) momentum maps, which show the momentum-resolved photoemission yield in the entire accessible momentum space range at a constant energy. These CE maps are the typical representation of the PT simulations since they directly reflect the periodicity of the molecular wave functions in real space.  

\begin{figure}[t]
	\centering
		\includegraphics[width=8.5cm]{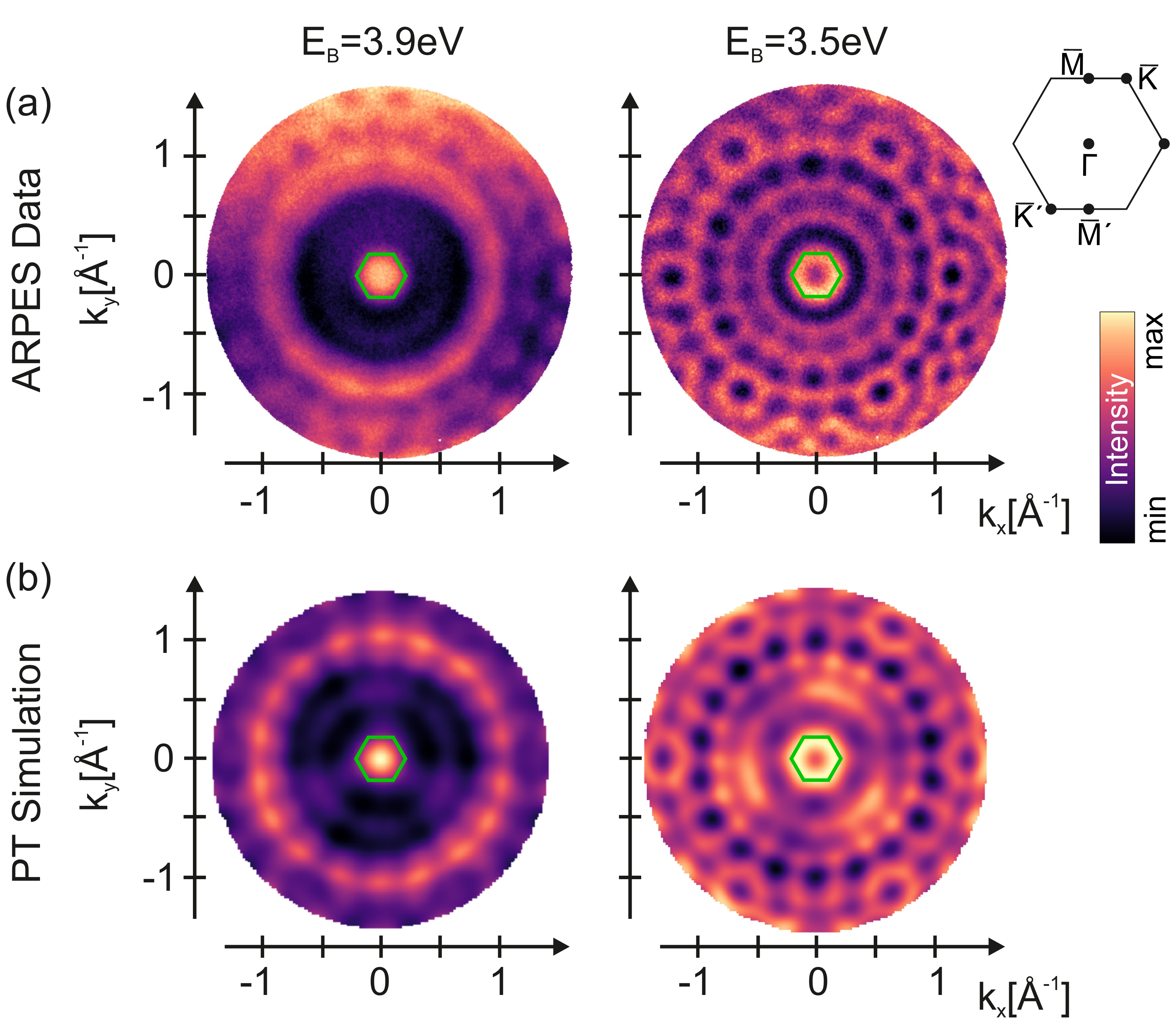}
	\caption{(a) Constant energy (CE) maps extracted from the ARPES data cube in the binding energy region of the HOMO-1 band at E$_\mathrm{B}=2.9\,$eV and E$_\mathrm{B}=3.5\,$eV. The corresponding CE maps predicted by our photoemission tomography simulation (PT) are shown in (b) for identical binding energies.} 
	\label{fig:Fig2}
\end{figure}

Fig.~2a shows CE maps extracted at two characteristic binding energies within the HOMO-1 band. The green hexagon marks the surface Brillouin zone of the C$_{60}$ crystal structure. All CE maps exhibit a quite complex momentum space pattern with sharp maxima which change their position and shape when scanning through the binding energy. For instance, the feature in the center of the surface Brillouin zone transforms from a dot-like emission at E$_\mathrm{B}=3.9\,$eV into a ring like emission at E$_\mathrm{B}=3.5\,$eV. This is clearly the spectroscopic signature of an upwards dispersing band in agreement with the energy vs. momentum cut in Fig.~1a. The emission features at larger momentum can be attributed to the same state repeated in the second and third Brillouin zones. 

The PT simulations at the corresponding binding energies within the HOMO-1 band are shown in Fig.~2b. These maps were obtained by considering the spectroscopic signatures of the main $(2\sqrt{3}\times 2\sqrt{3})R30^\circ$ structure as well as of the coexisting rotational domains as discussed in the supplementary material. The agreement between the PT simulations and our experimental data is striking and hence further confirms our previous assignment of the strongly dispersing bands to molecular orbitals with $\pi$-orbital character.

\begin{figure}[t!]
	\centering
		\includegraphics[width=8.5cm]{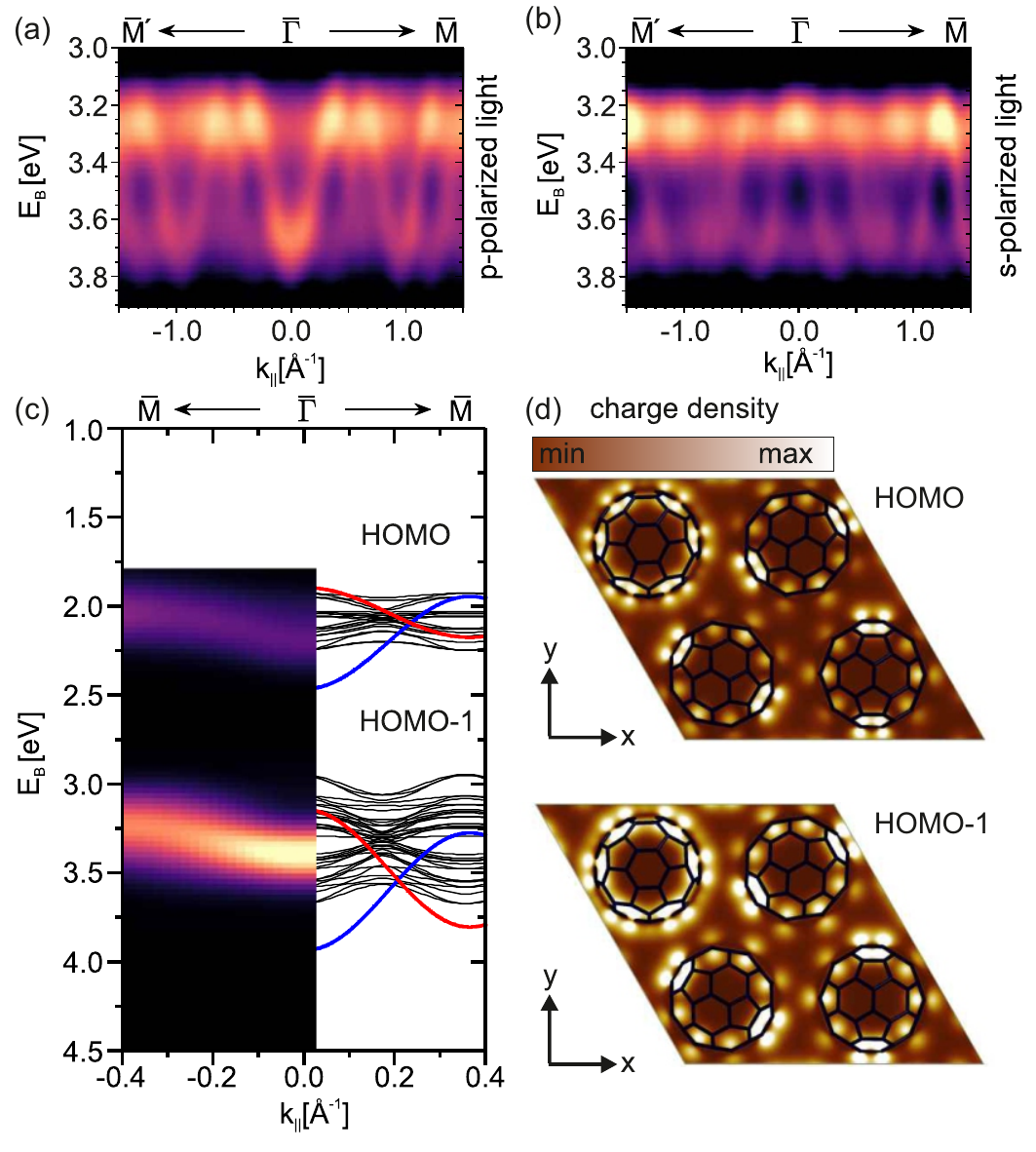}
	\caption{Energy vs. momentum cut of the HOMO-1 bands of a C$_{60}$ thin film recorded with vertical (a) and horizontal (b) light polarization (E$_\mathrm{photon}=35\,$eV). The contrast of both energy vs. momentum cuts is enhanced by using the second derivate of the experimental data. The left half of panel (c) shows the PT simulation of the energy vs. momentum cut in the first Brillouin zone, the right half the band structure of our density functional theory calculation. The red and blue solid lines are tight binding simulations to describe the experimental band dispersion obseverd for p- (blue) and s-poliarized light (red). (d) Real-space partial charge density distributions of the HOMO and HOMO-1 bands integrated in energy windows from $1.5$ to $2.5\,$eV and $2.75$ to $4.0\,$eV for the HOMO and HOMO-1 respectively. The partial charge density plots are shown in a top view of the C$_{60}$ unit cell in a plane through the center of the C$_{60}$ molecules.} 
	\label{fig:Fig3}
\end{figure}

To go beyond a pure qualitative analysis, we now focus on the band dispersion of the HOMO and HOMO-1 bands. The band dispersions of the HOMO-1 band were recorded with vertical (p) and horizontal (s) polarization of the synchrotron radiation (angle of incidence: $65^\circ$ with respect to the surface normal). The corresponding energy vs. momentum cuts are displayed in Fig.~3a and b, respectively. Both band dispersions exhibit clear qualitative differences depending on the light polarization. For p-polarized light, the HOMO-1 band at the $\bar{\Gamma}$-point disperses upward while it disperses downward for s-polarized light. This qualitative difference in the APRES data obtained with p- and s-polarized light demonstrates the existence of at least two bands with different orbital character in the binding energy range of the HOMO-1 state. Similar results were also observed for the HOMO band. 

For the quantification of the band dispersion, we extracted the diameter of the almost ring-like band in the first surface Brillouin zone for all CE maps of the HOMO and HOMO-1 bands (see supplementary material). For each orbital, we observe two bands, one with positive effective band mass (upwards dispersing band) and one with negative effective band mass (downwards dispersing band). The bands with positive effective mass are observed with p-polarized light suggesting a strong contribution of p$_{\mathrm{z}}$ orbitals of the C$_{60}$ thin film. In contrast, the bands with negative band mass are dominated by $\pi$-orbitals with a strong in-plane orbital character, i.e., with p$_{\mathrm{x/y}}$ orbitals character. The band dispersion is further analyzed by a tight-binding model calculated for a 2D hexagonal lattice \cite{Zhu.2006} which are shown in the right half of Fig.~3c as red and blue solid curves. The effective masses of the HOMO-1 bands are $\pm 5\,$m$_\mathrm{0}$, while the ones of the HOMO bands are $6\,$m$_\mathrm{0}$ and $- 10\,$m$_\mathrm{0}$, respectively. These band masses correspond to an oscillation bandwidth of $0.28$-$0.66\,$eV, in good agreement with previous studies of C$_{60}$ \cite{Tamai.2006,Latzke.2019,He.2006,Stadtmuller.2019}. 

The PT simulation of the band dispersion of the HOMO and HOMO-1 in the first surface Brillouin zone is shown in the left half of Fig.~3c. In contrast to our photoemission data, we only find one band with positive effective band mass, which is in qualitative agreement with our experimental findings obtained with p-polarized light. At this point, it is important to note that for the PT simulation, a plane-wave final state has been assumed. This approximation is known to work well for experimental geometries where the emission direction is close to parallel to the light polarization \cite{Puschnig.2009} corresponding to p-polarization in our case. Although the PT simulation can presently not account for the difference between p- and s-polarization, it clearly goes beyond a mere DFT band structure calculation. This is illustrated in the right half of Fig.~3c, which depicts all bands calculated with DFT for the unit cell containing four C$_{60}$ molecules: $4\times 5=20$ bands for the HOMO and $4 \times 9=36$ bands for the HOMO-1. Here, we observe both bands of positive and negative band mass in the binding energy range of the HOMO and HOMO-1 level, in agreement with a recent band structure calculation of a freestanding C$_{60}$ layer \cite{Latzke.2019}. By taking into account the photoemission cross sections, as is done in the PT simulation, only certain bands get selected which allows for a more realistic comparison with experimental ARPES data.

The large dispersion of the HOMO and HOMO-1 bands can also be understood by plotting the partial charge density distributions in Fig.~3d in a top view of the C$_{60}$ unit cell in a plane through the center of the C$_{60}$ molecules. For both orbitals, the charge density is not only localized on the molecular carbon cage but also in the free space between the fullerenes. The latter points to a significant overlap of the frontier molecular $\pi$-orbitals of neighboring molecules in the thin film which is hence responsible for the large band dispersion of the HOMO and HOMO-1 band of C$_{60}$. 

\begin{figure}[ht]
	\centering
		\includegraphics[width=8.5cm]{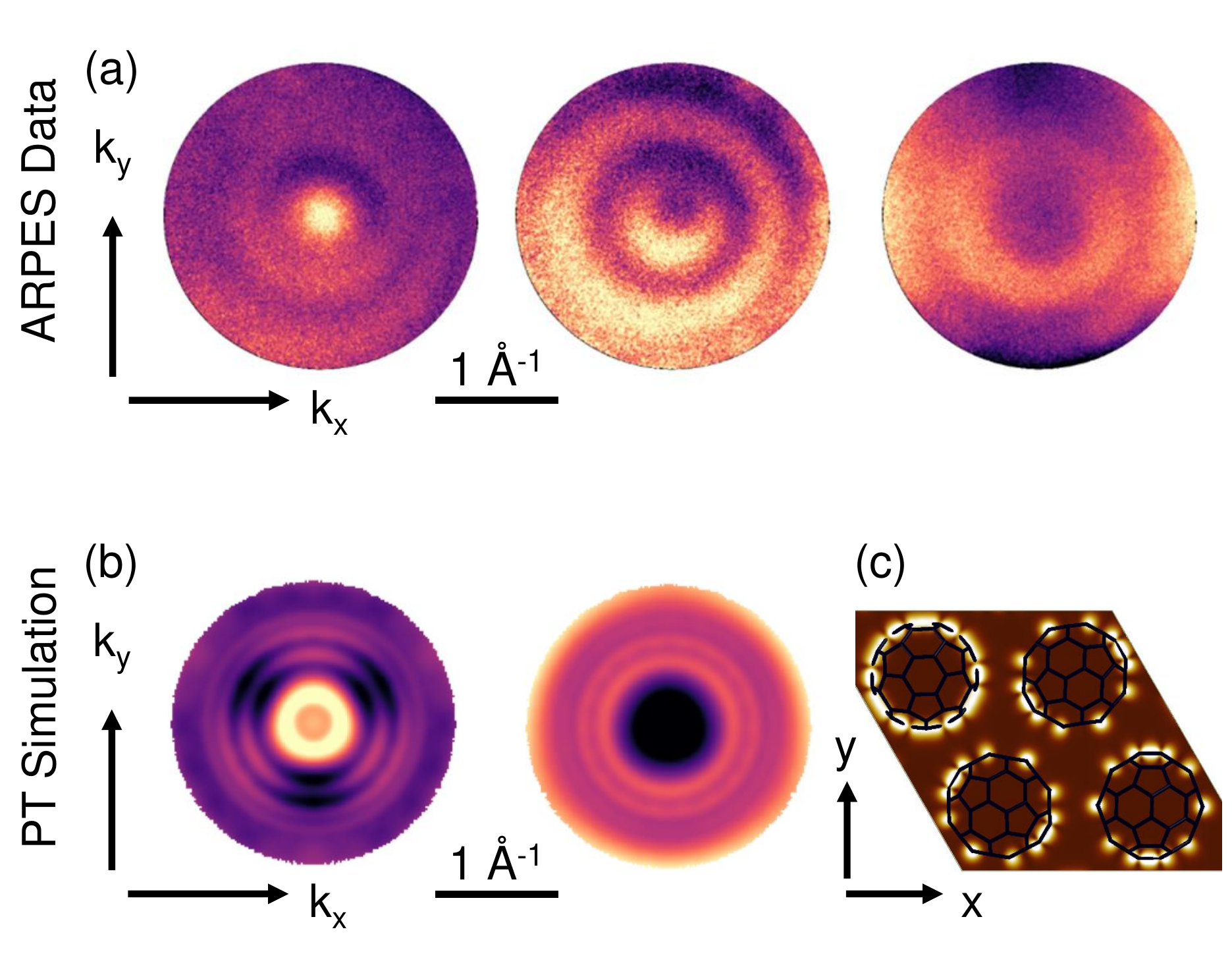}
	\caption{(a) Experimentally determined CE maps at three selected energies from left to right: $8.7\,$eV, $7.9\,$eV, and $7.4\,$eV. The corresponding PT simulations are shown in (b). Note that the CE maps in the PT simulations were extracted at slightly different binding energies as the experimental data. This is due to an energy difference in the initial state energy in our experiment and in the DFT calculations. (c) Partial charge density in real space of a $\sigma$-state integrated in energy windows from $6.0$ to $6.5\,$eV.} 
	\label{fig:Fig4}
\end{figure}

We now focus on the molecular states which can be found for E$_\mathrm{B}>5\,$eV. The photoemission signals in this binding energy range are dominated by $\sigma$-states with discrete maxima in the energy vs. momentum space. Interestingly, these discrete maxima are not randomly distributed in energy and momentum space but are all arranged along lines, see Fig.~1a. This observation is also directly reflected in the CE maps in Fig.~4a which were extracted at $8.7\,$eV, $7.9\,$eV, and $7.4\,$eV. They consist of ring-like intensity distributions with energy dependent radii which increase for smaller binding energies. Such a distinct correlation can be attributed to an intramolecular band dispersion of the molecular $\sigma$-states as observed by Koller et al. for crystalline films of sexiphenyl molecules \cite{Koller.2007}. The absence of a significant intermolecular dispersion of these bands can be understood by their $\sigma$-character which leads to a negligible intermolecular overlap of wave functions for adjacent molecules. 

Our experimental findings are qualitatively well reproduced by PT simulations. In particular, the simulated CE maps exhibit concentric emission features with increasing radius for smaller binding energies, see Fig.~4b. The overall emission pattern agree qualitatively with our ARPES data and enables us to gain insight into the spatial structure of the molecular orbitals in real space with high confidence. The spatial charge density distribution in Fig.~4c was integrated from $6.0$ to $6.5\,$eV. We find that the entire charge density is localized on the carbon cage of all four molecules of the unit cell while no charge density can be observed between the C$_{60}$ molecules. This clearly points to the absence of any overlap of the molecular wave functions for $\sigma$-states which explains the absence of intermolecular dispersion of these orbitals at large binding energies. 

Minor quantitative deviations between the ARPES data and the PT simulations can be attributed to the different initial state energy of the $\sigma$-states in the experiment and the band structure calculation in conjunction with the strong k$_\mathrm{z}$ dependency of the 3D Fourier transform of the localized molecular states of non-planar molecules. Both aspects are discussed in more detail in the supplementary material. 

In conclusion, we have provided new insights into the band dispersion and spatial delocalization of molecular orbitals of the prototypical three dimensional molecule C$_{60}$. Our photoemission experiment reveals two strongly dispersing molecular states with complex momentum-dependent photoemission patterns for small binding energies and non-dispersing emission maxima in energy and momentum space for larger binding energies. These different momentum-dependent photoemission distributions can be qualitatively described by our photoemission tomography simulations considering a plane wave final state, the inner potential of the C$_{60}$ thin film, and the band structure of a freestanding C$_{60}$ layer. This further confirms the applicability of PT to ARPES data of non-planar, three-dimensional molecular complexes \cite{Metzger.2020}. Even more importantly, it allows us to assign the strongly dispersing bands to molecular states with pure $\pi$-orbital character that are delocalized over neighbouring molecular sites and the non-dispersion emission pattern to localized $\sigma$-states of the individual C$_{60}$ molecules.

In this way, we were able to demonstrate the formation of an atomic crystal-like band structure in a molecular thin film. This is a vital step towards a yet unprecedented understanding of charge carrier transport in thin films of chemically designed molecules with superior functionalities, which, in most cases, exhibit a non-planar molecular structure.

\begin{acknowledgments}
The work was funded by the Deutsche Forschungsgemeinschaft (DFG, German Research Foundation) - TRR 173 - 268565370 (Project B05). B. S. and F.H. thankfully acknowledge financial support from the Graduate School of Excellence MAINZ (Excellence Initiative DFG/GSC 266). L.L.K. acknowledges financial support from Carl-Zeiss Stiftung for post-doctoral fellowship. M.C. and G.Z. acknowledge funding from the European Research Council (ERC) under the European Union's Horizon 2020 research and innovation programm (grant agreement no. 725767 - hyControl). D. L. and P. P. acknowledge support from the Austrian Science Fund (FWF) through project I3731. The computations have been performed on the HPC cluster of the KFU Graz and at the Vienna Scientific Computer (VSC). We acknowledge Elettra for providing the synchrotron radiation facility.
\end{acknowledgments}


\begin{thebibliography}{51}
\expandafter\ifx\csname natexlab\endcsname\relax\def\natexlab#1{#1}\fi
\expandafter\ifx\csname bibnamefont\endcsname\relax
  \def\bibnamefont#1{#1}\fi
\expandafter\ifx\csname bibfnamefont\endcsname\relax
  \def\bibfnamefont#1{#1}\fi
\expandafter\ifx\csname citenamefont\endcsname\relax
  \def\citenamefont#1{#1}\fi
\expandafter\ifx\csname url\endcsname\relax
  \def\url#1{\texttt{#1}}\fi
\expandafter\ifx\csname urlprefix\endcsname\relax\def\urlprefix{URL }\fi
\providecommand{\bibinfo}[2]{#2}
\providecommand{\eprint}[2][]{\url{#2}}

\bibitem[{\citenamefont{Schwarze et~al.}(2016)\citenamefont{Schwarze, Tress,
  Beyer, Gao, Scholz, Poelking, Ortstein, G{\"u}nther, Kasemann, Andrienko
  et~al.}}]{Schwarze.2016}
\bibinfo{author}{\bibfnamefont{M.}~\bibnamefont{Schwarze}},
  \bibinfo{author}{\bibfnamefont{W.}~\bibnamefont{Tress}},
  \bibinfo{author}{\bibfnamefont{B.}~\bibnamefont{Beyer}},
  \bibinfo{author}{\bibfnamefont{F.}~\bibnamefont{Gao}},
  \bibinfo{author}{\bibfnamefont{R.}~\bibnamefont{Scholz}},
  \bibinfo{author}{\bibfnamefont{C.}~\bibnamefont{Poelking}},
  \bibinfo{author}{\bibfnamefont{K.}~\bibnamefont{Ortstein}},
  \bibinfo{author}{\bibfnamefont{A.~A.} \bibnamefont{G{\"u}nther}},
  \bibinfo{author}{\bibfnamefont{D.}~\bibnamefont{Kasemann}},
  \bibinfo{author}{\bibfnamefont{D.}~\bibnamefont{Andrienko}},
  \bibnamefont{et~al.}, \bibinfo{journal}{Science}
  \textbf{\bibinfo{volume}{352}}, \bibinfo{pages}{1446} (\bibinfo{year}{2016}).

\bibitem[{\citenamefont{Bizzarri et~al.}(2017)\citenamefont{Bizzarri, Spuling,
  Knoll, Volz, and Br{\"a}se}}]{Bizzarri.2017}
\bibinfo{author}{\bibfnamefont{C.}~\bibnamefont{Bizzarri}},
  \bibinfo{author}{\bibfnamefont{E.}~\bibnamefont{Spuling}},
  \bibinfo{author}{\bibfnamefont{D.~M.} \bibnamefont{Knoll}},
  \bibinfo{author}{\bibfnamefont{D.}~\bibnamefont{Volz}}, \bibnamefont{and}
  \bibinfo{author}{\bibfnamefont{S.}~\bibnamefont{Br{\"a}se}},
  \bibinfo{journal}{Coord. Chem. Rev.}  (\bibinfo{year}{2017}).

\bibitem[{\citenamefont{Xu et~al.}(2014)\citenamefont{Xu, Chen, Sun, Lai, Su,
  Huang, and Liu}}]{Xu.2014}
\bibinfo{author}{\bibfnamefont{H.}~\bibnamefont{Xu}},
  \bibinfo{author}{\bibfnamefont{R.}~\bibnamefont{Chen}},
  \bibinfo{author}{\bibfnamefont{Q.}~\bibnamefont{Sun}},
  \bibinfo{author}{\bibfnamefont{W.}~\bibnamefont{Lai}},
  \bibinfo{author}{\bibfnamefont{Q.}~\bibnamefont{Su}},
  \bibinfo{author}{\bibfnamefont{W.}~\bibnamefont{Huang}}, \bibnamefont{and}
  \bibinfo{author}{\bibfnamefont{X.}~\bibnamefont{Liu}},
  \bibinfo{journal}{Chem. Soc. Rev.} \textbf{\bibinfo{volume}{43}},
  \bibinfo{pages}{3259} (\bibinfo{year}{2014}).

\bibitem[{\citenamefont{Cinchetti et~al.}(2017)\citenamefont{Cinchetti, Dediu,
  and Hueso}}]{Cinchetti.2017}
\bibinfo{author}{\bibfnamefont{M.}~\bibnamefont{Cinchetti}},
  \bibinfo{author}{\bibfnamefont{V.~A.} \bibnamefont{Dediu}}, \bibnamefont{and}
  \bibinfo{author}{\bibfnamefont{L.~E.} \bibnamefont{Hueso}},
  \bibinfo{journal}{Nat. Mater.} \textbf{\bibinfo{volume}{16}},
  \bibinfo{pages}{507} (\bibinfo{year}{2017}).

\bibitem[{\citenamefont{Anthony}(2006)}]{Anthony.2006}
\bibinfo{author}{\bibfnamefont{J.~E.} \bibnamefont{Anthony}},
  \bibinfo{journal}{Chem. Rev.} \textbf{\bibinfo{volume}{106}},
  \bibinfo{pages}{5028} (\bibinfo{year}{2006}).

\bibitem[{\citenamefont{Wang et~al.}(2012)\citenamefont{Wang, Dong, Hu, Liu,
  and Zhu}}]{Wang.2012}
\bibinfo{author}{\bibfnamefont{C.}~\bibnamefont{Wang}},
  \bibinfo{author}{\bibfnamefont{H.}~\bibnamefont{Dong}},
  \bibinfo{author}{\bibfnamefont{W.}~\bibnamefont{Hu}},
  \bibinfo{author}{\bibfnamefont{Y.}~\bibnamefont{Liu}}, \bibnamefont{and}
  \bibinfo{author}{\bibfnamefont{D.}~\bibnamefont{Zhu}},
  \bibinfo{journal}{Chem. Rev.} \textbf{\bibinfo{volume}{112}},
  \bibinfo{pages}{2208} (\bibinfo{year}{2012}).

\bibitem[{\citenamefont{Kang et~al.}(2013)\citenamefont{Kang, Yun, Chung, Kwon,
  and Kim}}]{Kang.2013}
\bibinfo{author}{\bibfnamefont{I.}~\bibnamefont{Kang}},
  \bibinfo{author}{\bibfnamefont{H.-J.} \bibnamefont{Yun}},
  \bibinfo{author}{\bibfnamefont{D.~S.} \bibnamefont{Chung}},
  \bibinfo{author}{\bibfnamefont{S.-K.} \bibnamefont{Kwon}}, \bibnamefont{and}
  \bibinfo{author}{\bibfnamefont{Y.-H.} \bibnamefont{Kim}},
  \bibinfo{journal}{J. Am. Chem. Soc} \textbf{\bibinfo{volume}{135}},
  \bibinfo{pages}{14896} (\bibinfo{year}{2013}).

\bibitem[{\citenamefont{Liu et~al.}(2015)\citenamefont{Liu, Zhang, Dong, Meng,
  Jiang, Jiang, Wang, Yu, Sun, Hu et~al.}}]{Liu.2015}
\bibinfo{author}{\bibfnamefont{J.}~\bibnamefont{Liu}},
  \bibinfo{author}{\bibfnamefont{H.}~\bibnamefont{Zhang}},
  \bibinfo{author}{\bibfnamefont{H.}~\bibnamefont{Dong}},
  \bibinfo{author}{\bibfnamefont{L.}~\bibnamefont{Meng}},
  \bibinfo{author}{\bibfnamefont{L.}~\bibnamefont{Jiang}},
  \bibinfo{author}{\bibfnamefont{L.}~\bibnamefont{Jiang}},
  \bibinfo{author}{\bibfnamefont{Y.}~\bibnamefont{Wang}},
  \bibinfo{author}{\bibfnamefont{J.}~\bibnamefont{Yu}},
  \bibinfo{author}{\bibfnamefont{Y.}~\bibnamefont{Sun}},
  \bibinfo{author}{\bibfnamefont{W.}~\bibnamefont{Hu}}, \bibnamefont{et~al.},
  \bibinfo{journal}{Nat. Commun.} \textbf{\bibinfo{volume}{6}},
  \bibinfo{pages}{10032} (\bibinfo{year}{2015}).

\bibitem[{\citenamefont{Ueno and Kera}(2008)}]{Ueno.2008}
\bibinfo{author}{\bibfnamefont{N.}~\bibnamefont{Ueno}} \bibnamefont{and}
  \bibinfo{author}{\bibfnamefont{S.}~\bibnamefont{Kera}},
  \bibinfo{journal}{Prog. Surf. Sci.} \textbf{\bibinfo{volume}{83}},
  \bibinfo{pages}{490} (\bibinfo{year}{2008}).

\bibitem[{\citenamefont{Br{\'e}das et~al.}(2004)\citenamefont{Br{\'e}das,
  Beljonne, Coropceanu, and Cornil}}]{Bredas.2004}
\bibinfo{author}{\bibfnamefont{J.-L.} \bibnamefont{Br{\'e}das}},
  \bibinfo{author}{\bibfnamefont{D.}~\bibnamefont{Beljonne}},
  \bibinfo{author}{\bibfnamefont{V.}~\bibnamefont{Coropceanu}},
  \bibnamefont{and} \bibinfo{author}{\bibfnamefont{J.}~\bibnamefont{Cornil}},
  \bibinfo{journal}{Chem. Rev.} \textbf{\bibinfo{volume}{104}},
  \bibinfo{pages}{4971} (\bibinfo{year}{2004}).

\bibitem[{\citenamefont{Kera and Ueno}(2015)}]{Kera.2015}
\bibinfo{author}{\bibfnamefont{S.}~\bibnamefont{Kera}} \bibnamefont{and}
  \bibinfo{author}{\bibfnamefont{N.}~\bibnamefont{Ueno}}, \bibinfo{journal}{J.
  Electron. Spectrosc. Rel. Phenom.} \textbf{\bibinfo{volume}{204}},
  \bibinfo{pages}{2} (\bibinfo{year}{2015}).

\bibitem[{\citenamefont{Machida et~al.}(2010)\citenamefont{Machida, Nakayama,
  Duhm, Xin, Funakoshi, Ogawa, Kera, Ueno, and Ishii}}]{Machida.2010}
\bibinfo{author}{\bibfnamefont{S.-i.} \bibnamefont{Machida}},
  \bibinfo{author}{\bibfnamefont{Y.}~\bibnamefont{Nakayama}},
  \bibinfo{author}{\bibfnamefont{S.}~\bibnamefont{Duhm}},
  \bibinfo{author}{\bibfnamefont{Q.}~\bibnamefont{Xin}},
  \bibinfo{author}{\bibfnamefont{A.}~\bibnamefont{Funakoshi}},
  \bibinfo{author}{\bibfnamefont{N.}~\bibnamefont{Ogawa}},
  \bibinfo{author}{\bibfnamefont{S.}~\bibnamefont{Kera}},
  \bibinfo{author}{\bibfnamefont{N.}~\bibnamefont{Ueno}}, \bibnamefont{and}
  \bibinfo{author}{\bibfnamefont{H.}~\bibnamefont{Ishii}},
  \bibinfo{journal}{Phys. Rev. Lett.} \textbf{\bibinfo{volume}{104}},
  \bibinfo{pages}{156401} (\bibinfo{year}{2010}).

\bibitem[{\citenamefont{Bussolotti et~al.}(2017)\citenamefont{Bussolotti, Yang,
  Yamaguchi, Yonezawa, Sato, Matsunami, Tanaka, Nakayama, Ishii, Ueno
  et~al.}}]{Bussolotti.2017}
\bibinfo{author}{\bibfnamefont{F.}~\bibnamefont{Bussolotti}},
  \bibinfo{author}{\bibfnamefont{J.}~\bibnamefont{Yang}},
  \bibinfo{author}{\bibfnamefont{T.}~\bibnamefont{Yamaguchi}},
  \bibinfo{author}{\bibfnamefont{K.}~\bibnamefont{Yonezawa}},
  \bibinfo{author}{\bibfnamefont{K.}~\bibnamefont{Sato}},
  \bibinfo{author}{\bibfnamefont{M.}~\bibnamefont{Matsunami}},
  \bibinfo{author}{\bibfnamefont{K.}~\bibnamefont{Tanaka}},
  \bibinfo{author}{\bibfnamefont{Y.}~\bibnamefont{Nakayama}},
  \bibinfo{author}{\bibfnamefont{H.}~\bibnamefont{Ishii}},
  \bibinfo{author}{\bibfnamefont{N.}~\bibnamefont{Ueno}}, \bibnamefont{et~al.},
  \bibinfo{journal}{Nat. Commun.} \textbf{\bibinfo{volume}{8}},
  \bibinfo{pages}{173} (\bibinfo{year}{2017}).

\bibitem[{\citenamefont{Latzke et~al.}(2019)\citenamefont{Latzke,
  Ojeda-Aristizabal, Griffin, Denlinger, Neaton, Zettl, and
  Lanzara}}]{Latzke.2019}
\bibinfo{author}{\bibfnamefont{D.~W.} \bibnamefont{Latzke}},
  \bibinfo{author}{\bibfnamefont{C.}~\bibnamefont{Ojeda-Aristizabal}},
  \bibinfo{author}{\bibfnamefont{S.~M.} \bibnamefont{Griffin}},
  \bibinfo{author}{\bibfnamefont{J.~D.} \bibnamefont{Denlinger}},
  \bibinfo{author}{\bibfnamefont{J.~B.} \bibnamefont{Neaton}},
  \bibinfo{author}{\bibfnamefont{A.}~\bibnamefont{Zettl}}, \bibnamefont{and}
  \bibinfo{author}{\bibfnamefont{A.}~\bibnamefont{Lanzara}},
  \bibinfo{journal}{Phys. Rev. B} \textbf{\bibinfo{volume}{99}},
  \bibinfo{pages}{045425} (\bibinfo{year}{2019}).

\bibitem[{\citenamefont{Menard et~al.}(2004)\citenamefont{Menard, Podzorov,
  Hur, Gaur, Gershenson, and Rogers}}]{Menard.2004}
\bibinfo{author}{\bibfnamefont{E.}~\bibnamefont{Menard}},
  \bibinfo{author}{\bibfnamefont{V.}~\bibnamefont{Podzorov}},
  \bibinfo{author}{\bibfnamefont{S.-H.} \bibnamefont{Hur}},
  \bibinfo{author}{\bibfnamefont{A.}~\bibnamefont{Gaur}},
  \bibinfo{author}{\bibfnamefont{M.~E.} \bibnamefont{Gershenson}},
  \bibnamefont{and} \bibinfo{author}{\bibfnamefont{J.~A.}
  \bibnamefont{Rogers}}, \bibinfo{journal}{Adv. Mater.}
  \textbf{\bibinfo{volume}{16}}, \bibinfo{pages}{2097} (\bibinfo{year}{2004}).

\bibitem[{\citenamefont{Sundar et~al.}(2004)\citenamefont{Sundar, Zaumseil,
  Podzorov, Menard, Willett, Someya, Gershenson, and Rogers}}]{Sundar.2004}
\bibinfo{author}{\bibfnamefont{V.~C.} \bibnamefont{Sundar}},
  \bibinfo{author}{\bibfnamefont{J.}~\bibnamefont{Zaumseil}},
  \bibinfo{author}{\bibfnamefont{V.}~\bibnamefont{Podzorov}},
  \bibinfo{author}{\bibfnamefont{E.}~\bibnamefont{Menard}},
  \bibinfo{author}{\bibfnamefont{R.~L.} \bibnamefont{Willett}},
  \bibinfo{author}{\bibfnamefont{T.}~\bibnamefont{Someya}},
  \bibinfo{author}{\bibfnamefont{M.~E.} \bibnamefont{Gershenson}},
  \bibnamefont{and} \bibinfo{author}{\bibfnamefont{J.~A.}
  \bibnamefont{Rogers}}, \bibinfo{journal}{Science}
  \textbf{\bibinfo{volume}{303}}, \bibinfo{pages}{1644} (\bibinfo{year}{2004}).

\bibitem[{\citenamefont{Anthopoulos et~al.}(2004)\citenamefont{Anthopoulos,
  Tanase, Setayesh, Meijer, Hummelen, Blom, and de~Leeuw}}]{Anthopoulos.2004}
\bibinfo{author}{\bibfnamefont{T.~D.} \bibnamefont{Anthopoulos}},
  \bibinfo{author}{\bibfnamefont{C.}~\bibnamefont{Tanase}},
  \bibinfo{author}{\bibfnamefont{S.}~\bibnamefont{Setayesh}},
  \bibinfo{author}{\bibfnamefont{E.~J.} \bibnamefont{Meijer}},
  \bibinfo{author}{\bibfnamefont{J.~C.} \bibnamefont{Hummelen}},
  \bibinfo{author}{\bibfnamefont{P.~W.~M.} \bibnamefont{Blom}},
  \bibnamefont{and} \bibinfo{author}{\bibfnamefont{D.~M.}
  \bibnamefont{de~Leeuw}}, \bibinfo{journal}{Adv. Mater.}
  \textbf{\bibinfo{volume}{16}}, \bibinfo{pages}{2174} (\bibinfo{year}{2004}).

\bibitem[{\citenamefont{Puschnig et~al.}(2009)\citenamefont{Puschnig,
  Berkebile, Fleming, Koller, Emtsev, Seyller, Riley, Ambrosch-Draxl, Netzer,
  and Ramsey}}]{Puschnig.2009}
\bibinfo{author}{\bibfnamefont{P.}~\bibnamefont{Puschnig}},
  \bibinfo{author}{\bibfnamefont{S.}~\bibnamefont{Berkebile}},
  \bibinfo{author}{\bibfnamefont{A.~J.} \bibnamefont{Fleming}},
  \bibinfo{author}{\bibfnamefont{G.}~\bibnamefont{Koller}},
  \bibinfo{author}{\bibfnamefont{K.}~\bibnamefont{Emtsev}},
  \bibinfo{author}{\bibfnamefont{T.}~\bibnamefont{Seyller}},
  \bibinfo{author}{\bibfnamefont{J.~D.} \bibnamefont{Riley}},
  \bibinfo{author}{\bibfnamefont{C.}~\bibnamefont{Ambrosch-Draxl}},
  \bibinfo{author}{\bibfnamefont{F.~P.} \bibnamefont{Netzer}},
  \bibnamefont{and} \bibinfo{author}{\bibfnamefont{M.~G.}
  \bibnamefont{Ramsey}}, \bibinfo{journal}{Science}
  \textbf{\bibinfo{volume}{326}}, \bibinfo{pages}{702} (\bibinfo{year}{2009}).

\bibitem[{\citenamefont{Ziroff et~al.}(2010)\citenamefont{Ziroff, Forster,
  Sch{\"o}ll, Puschnig, and Reinert}}]{Ziroff.2010}
\bibinfo{author}{\bibfnamefont{J.}~\bibnamefont{Ziroff}},
  \bibinfo{author}{\bibfnamefont{F.}~\bibnamefont{Forster}},
  \bibinfo{author}{\bibfnamefont{A.}~\bibnamefont{Sch{\"o}ll}},
  \bibinfo{author}{\bibfnamefont{P.}~\bibnamefont{Puschnig}}, \bibnamefont{and}
  \bibinfo{author}{\bibfnamefont{F.}~\bibnamefont{Reinert}},
  \bibinfo{journal}{Phys. Rev. Lett.} \textbf{\bibinfo{volume}{104}},
  \bibinfo{pages}{233004} (\bibinfo{year}{2010}).

\bibitem[{\citenamefont{Willenbockel et~al.}(2013)\citenamefont{Willenbockel,
  Stadtm{\"u}ller, Sch{\"o}nauer, Bocquet, L{\"u}ftner, Reinisch, Ules, Koller,
  Kumpf, Soubatch et~al.}}]{Willenbockel.2013}
\bibinfo{author}{\bibfnamefont{M.}~\bibnamefont{Willenbockel}},
  \bibinfo{author}{\bibfnamefont{B.}~\bibnamefont{Stadtm{\"u}ller}},
  \bibinfo{author}{\bibfnamefont{K.}~\bibnamefont{Sch{\"o}nauer}},
  \bibinfo{author}{\bibfnamefont{F.~C.} \bibnamefont{Bocquet}},
  \bibinfo{author}{\bibfnamefont{D.}~\bibnamefont{L{\"u}ftner}},
  \bibinfo{author}{\bibfnamefont{E.~M.} \bibnamefont{Reinisch}},
  \bibinfo{author}{\bibfnamefont{T.}~\bibnamefont{Ules}},
  \bibinfo{author}{\bibfnamefont{G.}~\bibnamefont{Koller}},
  \bibinfo{author}{\bibfnamefont{C.}~\bibnamefont{Kumpf}},
  \bibinfo{author}{\bibfnamefont{S.}~\bibnamefont{Soubatch}},
  \bibnamefont{et~al.}, \bibinfo{journal}{New J. Phys.}
  \textbf{\bibinfo{volume}{15}}, \bibinfo{pages}{033017}
  (\bibinfo{year}{2013}).

\bibitem[{\citenamefont{Feyer et~al.}(2014)\citenamefont{Feyer, Graus, Nigge,
  Wie{\ss}ner, Acres, Wiemann, Schneider, Sch{\"o}ll, and
  Reinert}}]{Feyer.2014}
\bibinfo{author}{\bibfnamefont{V.}~\bibnamefont{Feyer}},
  \bibinfo{author}{\bibfnamefont{M.}~\bibnamefont{Graus}},
  \bibinfo{author}{\bibfnamefont{P.}~\bibnamefont{Nigge}},
  \bibinfo{author}{\bibfnamefont{M.}~\bibnamefont{Wie{\ss}ner}},
  \bibinfo{author}{\bibfnamefont{R.~G.} \bibnamefont{Acres}},
  \bibinfo{author}{\bibfnamefont{C.}~\bibnamefont{Wiemann}},
  \bibinfo{author}{\bibfnamefont{C.~M.} \bibnamefont{Schneider}},
  \bibinfo{author}{\bibfnamefont{A.}~\bibnamefont{Sch{\"o}ll}},
  \bibnamefont{and} \bibinfo{author}{\bibfnamefont{F.}~\bibnamefont{Reinert}},
  \bibinfo{journal}{Surf. Sci.} \textbf{\bibinfo{volume}{621}},
  \bibinfo{pages}{64} (\bibinfo{year}{2014}).

\bibitem[{\citenamefont{Grimm et~al.}(2018)\citenamefont{Grimm, Metzger, Graus,
  Jugovac, Zamborlini, Feyer, Sch{\"o}ll, and Reinert}}]{Grimm.2018}
\bibinfo{author}{\bibfnamefont{M.}~\bibnamefont{Grimm}},
  \bibinfo{author}{\bibfnamefont{C.}~\bibnamefont{Metzger}},
  \bibinfo{author}{\bibfnamefont{M.}~\bibnamefont{Graus}},
  \bibinfo{author}{\bibfnamefont{M.}~\bibnamefont{Jugovac}},
  \bibinfo{author}{\bibfnamefont{G.}~\bibnamefont{Zamborlini}},
  \bibinfo{author}{\bibfnamefont{V.}~\bibnamefont{Feyer}},
  \bibinfo{author}{\bibfnamefont{A.}~\bibnamefont{Sch{\"o}ll}},
  \bibnamefont{and} \bibinfo{author}{\bibfnamefont{F.}~\bibnamefont{Reinert}},
  \bibinfo{journal}{Phys. Rev. B} \textbf{\bibinfo{volume}{98}},
  \bibinfo{pages}{195412} (\bibinfo{year}{2018}).

\bibitem[{\citenamefont{Stadtm{\"u}ller
  et~al.}(2012)\citenamefont{Stadtm{\"u}ller, Willenbockel, Reinisch, Ules,
  Bocquet, Soubatch, Puschnig, Koller, Ramsey, Tautz
  et~al.}}]{Stadtmuller.2012}
\bibinfo{author}{\bibfnamefont{B.}~\bibnamefont{Stadtm{\"u}ller}},
  \bibinfo{author}{\bibfnamefont{M.}~\bibnamefont{Willenbockel}},
  \bibinfo{author}{\bibfnamefont{E.~M.} \bibnamefont{Reinisch}},
  \bibinfo{author}{\bibfnamefont{T.}~\bibnamefont{Ules}},
  \bibinfo{author}{\bibfnamefont{F.~C.} \bibnamefont{Bocquet}},
  \bibinfo{author}{\bibfnamefont{S.}~\bibnamefont{Soubatch}},
  \bibinfo{author}{\bibfnamefont{P.}~\bibnamefont{Puschnig}},
  \bibinfo{author}{\bibfnamefont{G.}~\bibnamefont{Koller}},
  \bibinfo{author}{\bibfnamefont{M.~G.} \bibnamefont{Ramsey}},
  \bibinfo{author}{\bibfnamefont{F.~S.} \bibnamefont{Tautz}},
  \bibnamefont{et~al.}, \bibinfo{journal}{EPL} \textbf{\bibinfo{volume}{100}},
  \bibinfo{pages}{26008} (\bibinfo{year}{2012}).

\bibitem[{\citenamefont{Weiss et~al.}(2015)\citenamefont{Weiss, Luftner, Ules,
  Reinisch, Kaser, Gottwald, Richter, Soubatch, Koller, Ramsey
  et~al.}}]{Weiss.2015}
\bibinfo{author}{\bibfnamefont{S.}~\bibnamefont{Weiss}},
  \bibinfo{author}{\bibfnamefont{D.}~\bibnamefont{Luftner}},
  \bibinfo{author}{\bibfnamefont{T.}~\bibnamefont{Ules}},
  \bibinfo{author}{\bibfnamefont{E.~M.} \bibnamefont{Reinisch}},
  \bibinfo{author}{\bibfnamefont{H.}~\bibnamefont{Kaser}},
  \bibinfo{author}{\bibfnamefont{A.}~\bibnamefont{Gottwald}},
  \bibinfo{author}{\bibfnamefont{M.}~\bibnamefont{Richter}},
  \bibinfo{author}{\bibfnamefont{S.}~\bibnamefont{Soubatch}},
  \bibinfo{author}{\bibfnamefont{G.}~\bibnamefont{Koller}},
  \bibinfo{author}{\bibfnamefont{M.~G.} \bibnamefont{Ramsey}},
  \bibnamefont{et~al.}, \bibinfo{journal}{Nat. Commun.}
  \textbf{\bibinfo{volume}{6}}, \bibinfo{pages}{8287} (\bibinfo{year}{2015}).

\bibitem[{\citenamefont{Egger et~al.}(2019)\citenamefont{Egger, Kollmann,
  Hurdax, L{\"u}ftner, Yang, Weiss, Gottwald, Richter, Koller, Soubatch
  et~al.}}]{Egger.2019}
\bibinfo{author}{\bibfnamefont{L.}~\bibnamefont{Egger}},
  \bibinfo{author}{\bibfnamefont{B.}~\bibnamefont{Kollmann}},
  \bibinfo{author}{\bibfnamefont{P.}~\bibnamefont{Hurdax}},
  \bibinfo{author}{\bibfnamefont{D.}~\bibnamefont{L{\"u}ftner}},
  \bibinfo{author}{\bibfnamefont{X.}~\bibnamefont{Yang}},
  \bibinfo{author}{\bibfnamefont{S.}~\bibnamefont{Weiss}},
  \bibinfo{author}{\bibfnamefont{A.}~\bibnamefont{Gottwald}},
  \bibinfo{author}{\bibfnamefont{M.}~\bibnamefont{Richter}},
  \bibinfo{author}{\bibfnamefont{G.}~\bibnamefont{Koller}},
  \bibinfo{author}{\bibfnamefont{S.}~\bibnamefont{Soubatch}},
  \bibnamefont{et~al.}, \bibinfo{journal}{New J. Phys.}
  \textbf{\bibinfo{volume}{21}}, \bibinfo{pages}{043003}
  (\bibinfo{year}{2019}).

\bibitem[{\citenamefont{Zamborlini et~al.}(2017)\citenamefont{Zamborlini,
  L{\"u}ftner, Feng, Kollmann, Puschnig, Dri, Panighel, {Di Santo}, Goldoni,
  Comelli et~al.}}]{Zamborlini.2017}
\bibinfo{author}{\bibfnamefont{G.}~\bibnamefont{Zamborlini}},
  \bibinfo{author}{\bibfnamefont{D.}~\bibnamefont{L{\"u}ftner}},
  \bibinfo{author}{\bibfnamefont{Z.}~\bibnamefont{Feng}},
  \bibinfo{author}{\bibfnamefont{B.}~\bibnamefont{Kollmann}},
  \bibinfo{author}{\bibfnamefont{P.}~\bibnamefont{Puschnig}},
  \bibinfo{author}{\bibfnamefont{C.}~\bibnamefont{Dri}},
  \bibinfo{author}{\bibfnamefont{M.}~\bibnamefont{Panighel}},
  \bibinfo{author}{\bibfnamefont{G.}~\bibnamefont{{Di Santo}}},
  \bibinfo{author}{\bibfnamefont{A.}~\bibnamefont{Goldoni}},
  \bibinfo{author}{\bibfnamefont{G.}~\bibnamefont{Comelli}},
  \bibnamefont{et~al.}, \bibinfo{journal}{Nat. Commun.}
  \textbf{\bibinfo{volume}{8}}, \bibinfo{pages}{335} (\bibinfo{year}{2017}).

\bibitem[{\citenamefont{Stadtm{\"u}ller
  et~al.}(2014)\citenamefont{Stadtm{\"u}ller, L{\"u}ftner, Willenbockel,
  Reinisch, Sueyoshi, Koller, Soubatch, Ramsey, Puschnig, Tautz
  et~al.}}]{Stadtmuller.2014}
\bibinfo{author}{\bibfnamefont{B.}~\bibnamefont{Stadtm{\"u}ller}},
  \bibinfo{author}{\bibfnamefont{D.}~\bibnamefont{L{\"u}ftner}},
  \bibinfo{author}{\bibfnamefont{M.}~\bibnamefont{Willenbockel}},
  \bibinfo{author}{\bibfnamefont{E.~M.} \bibnamefont{Reinisch}},
  \bibinfo{author}{\bibfnamefont{T.}~\bibnamefont{Sueyoshi}},
  \bibinfo{author}{\bibfnamefont{G.}~\bibnamefont{Koller}},
  \bibinfo{author}{\bibfnamefont{S.}~\bibnamefont{Soubatch}},
  \bibinfo{author}{\bibfnamefont{M.~G.} \bibnamefont{Ramsey}},
  \bibinfo{author}{\bibfnamefont{P.}~\bibnamefont{Puschnig}},
  \bibinfo{author}{\bibfnamefont{F.~S.} \bibnamefont{Tautz}},
  \bibnamefont{et~al.}, \bibinfo{journal}{Nat. Commun.}
  \textbf{\bibinfo{volume}{5}}, \bibinfo{pages}{3685} (\bibinfo{year}{2014}).

\bibitem[{\citenamefont{Willenbockel et~al.}(2015)\citenamefont{Willenbockel,
  L{\"u}ftner, Stadtm{\"u}ller, Koller, Kumpf, Soubatch, Puschnig, Ramsey, and
  Tautz}}]{Willenbockel.2015}
\bibinfo{author}{\bibfnamefont{M.}~\bibnamefont{Willenbockel}},
  \bibinfo{author}{\bibfnamefont{D.}~\bibnamefont{L{\"u}ftner}},
  \bibinfo{author}{\bibfnamefont{B.}~\bibnamefont{Stadtm{\"u}ller}},
  \bibinfo{author}{\bibfnamefont{G.}~\bibnamefont{Koller}},
  \bibinfo{author}{\bibfnamefont{C.}~\bibnamefont{Kumpf}},
  \bibinfo{author}{\bibfnamefont{S.}~\bibnamefont{Soubatch}},
  \bibinfo{author}{\bibfnamefont{P.}~\bibnamefont{Puschnig}},
  \bibinfo{author}{\bibfnamefont{M.~G.} \bibnamefont{Ramsey}},
  \bibnamefont{and} \bibinfo{author}{\bibfnamefont{F.~S.} \bibnamefont{Tautz}},
  \bibinfo{journal}{Phys. Chem. Chem. Phys.} \textbf{\bibinfo{volume}{17}},
  \bibinfo{pages}{1530} (\bibinfo{year}{2015}).

\bibitem[{\citenamefont{Metzger et~al.}(2020)\citenamefont{Metzger, Graus,
  Grimm, Zamborlini, Feyer, Schwendt, L{\"u}ftner, Puschnig, Sch{\"o}ll, and
  Reinert}}]{Metzger.2020}
\bibinfo{author}{\bibfnamefont{C.}~\bibnamefont{Metzger}},
  \bibinfo{author}{\bibfnamefont{M.}~\bibnamefont{Graus}},
  \bibinfo{author}{\bibfnamefont{M.}~\bibnamefont{Grimm}},
  \bibinfo{author}{\bibfnamefont{G.}~\bibnamefont{Zamborlini}},
  \bibinfo{author}{\bibfnamefont{V.}~\bibnamefont{Feyer}},
  \bibinfo{author}{\bibfnamefont{M.}~\bibnamefont{Schwendt}},
  \bibinfo{author}{\bibfnamefont{D.}~\bibnamefont{L{\"u}ftner}},
  \bibinfo{author}{\bibfnamefont{P.}~\bibnamefont{Puschnig}},
  \bibinfo{author}{\bibfnamefont{A.}~\bibnamefont{Sch{\"o}ll}},
  \bibnamefont{and} \bibinfo{author}{\bibfnamefont{F.}~\bibnamefont{Reinert}},
  \bibinfo{journal}{{Phys. Rev. B}} \textbf{\bibinfo{volume}{101}},
  \bibinfo{pages}{165421} (\bibinfo{year}{2020}).

\bibitem[{\citenamefont{Dresselhaus et~al.}(1996)\citenamefont{Dresselhaus,
  Dresselhaus, and Eklund}}]{Dresselhaus.1996}
\bibinfo{author}{\bibfnamefont{M.~S.} \bibnamefont{Dresselhaus}},
  \bibinfo{author}{\bibfnamefont{G.}~\bibnamefont{Dresselhaus}},
  \bibnamefont{and} \bibinfo{author}{\bibfnamefont{P.~C.}
  \bibnamefont{Eklund}}, \bibinfo{title}{Science of Fullerenes and Carbon
  Nanotubes: Their Properties and Applications} (\bibinfo{publisher}{{Elsevier
  professional}}, \bibinfo{address}{s.l.}, \bibinfo{year}{1996}),
  \bibinfo{edition}{1st} ed., ISBN \bibinfo{isbn}{0122218205}.

\bibitem[{\citenamefont{David et~al.}(1992)\citenamefont{David, Ibberson,
  Dennis, Hare, and Prassides}}]{David.1992}
\bibinfo{author}{\bibfnamefont{W.~I.~F.} \bibnamefont{David}},
  \bibinfo{author}{\bibfnamefont{R.~M.} \bibnamefont{Ibberson}},
  \bibinfo{author}{\bibfnamefont{T.~J.~S.} \bibnamefont{Dennis}},
  \bibinfo{author}{\bibfnamefont{J.~P.} \bibnamefont{Hare}}, \bibnamefont{and}
  \bibinfo{author}{\bibfnamefont{K.}~\bibnamefont{Prassides}},
  \bibinfo{journal}{EPL} \textbf{\bibinfo{volume}{18}}, \bibinfo{pages}{219}
  (\bibinfo{year}{1992}).

\bibitem[{\citenamefont{David et~al.}(1991)\citenamefont{David, Ibberson,
  Matthewman, Prassides, Dennis, Hare, Kroto, Taylor, and Walton}}]{David.1991}
\bibinfo{author}{\bibfnamefont{W.~I.~F.} \bibnamefont{David}},
  \bibinfo{author}{\bibfnamefont{R.~M.} \bibnamefont{Ibberson}},
  \bibinfo{author}{\bibfnamefont{J.~C.} \bibnamefont{Matthewman}},
  \bibinfo{author}{\bibfnamefont{K.}~\bibnamefont{Prassides}},
  \bibinfo{author}{\bibfnamefont{T.~J.~S.} \bibnamefont{Dennis}},
  \bibinfo{author}{\bibfnamefont{J.~P.} \bibnamefont{Hare}},
  \bibinfo{author}{\bibfnamefont{H.~W.} \bibnamefont{Kroto}},
  \bibinfo{author}{\bibfnamefont{R.}~\bibnamefont{Taylor}}, \bibnamefont{and}
  \bibinfo{author}{\bibfnamefont{D.~R.~M.} \bibnamefont{Walton}},
  \bibinfo{journal}{Nature} \textbf{\bibinfo{volume}{353}},
  \bibinfo{pages}{147} (\bibinfo{year}{1991}).

\bibitem[{\citenamefont{Kr{\"o}mker et~al.}(2008)\citenamefont{Kr{\"o}mker,
  Escher, Funnemann, Hartung, Engelhard, and Kirschner}}]{Kromker.2008}
\bibinfo{author}{\bibfnamefont{B.}~\bibnamefont{Kr{\"o}mker}},
  \bibinfo{author}{\bibfnamefont{M.}~\bibnamefont{Escher}},
  \bibinfo{author}{\bibfnamefont{D.}~\bibnamefont{Funnemann}},
  \bibinfo{author}{\bibfnamefont{D.}~\bibnamefont{Hartung}},
  \bibinfo{author}{\bibfnamefont{H.}~\bibnamefont{Engelhard}},
  \bibnamefont{and}
  \bibinfo{author}{\bibfnamefont{J.}~\bibnamefont{Kirschner}},
  \bibinfo{journal}{Rev. Sci. Instrum.} \textbf{\bibinfo{volume}{79}},
  \bibinfo{pages}{053702} (\bibinfo{year}{2008}).

\bibitem[{\citenamefont{Tusche et~al.}(2013)\citenamefont{Tusche, Ellguth,
  Krasyuk, Winkelmann, Kutnyakhov, Lushchyk, Medjanik, Schonhense, and
  Kirschner}}]{Tusche.2013}
\bibinfo{author}{\bibfnamefont{C.}~\bibnamefont{Tusche}},
  \bibinfo{author}{\bibfnamefont{M.}~\bibnamefont{Ellguth}},
  \bibinfo{author}{\bibfnamefont{A.}~\bibnamefont{Krasyuk}},
  \bibinfo{author}{\bibfnamefont{A.}~\bibnamefont{Winkelmann}},
  \bibinfo{author}{\bibfnamefont{D.}~\bibnamefont{Kutnyakhov}},
  \bibinfo{author}{\bibfnamefont{P.}~\bibnamefont{Lushchyk}},
  \bibinfo{author}{\bibfnamefont{K.}~\bibnamefont{Medjanik}},
  \bibinfo{author}{\bibfnamefont{G.}~\bibnamefont{Schonhense}},
  \bibnamefont{and}
  \bibinfo{author}{\bibfnamefont{J.}~\bibnamefont{Kirschner}},
  \bibinfo{journal}{Ultramicroscopy} \textbf{\bibinfo{volume}{130}},
  \bibinfo{pages}{70} (\bibinfo{year}{2013}).

\bibitem[{\citenamefont{Shi et~al.}(2012)\citenamefont{Shi, {van Hove}, and
  Zhang}}]{Shi.2012}
\bibinfo{author}{\bibfnamefont{X.-Q.} \bibnamefont{Shi}},
  \bibinfo{author}{\bibfnamefont{M.~A.} \bibnamefont{{van Hove}}},
  \bibnamefont{and} \bibinfo{author}{\bibfnamefont{R.-Q.} \bibnamefont{Zhang}},
  \bibinfo{journal}{J. Mater. Sci.} \textbf{\bibinfo{volume}{47}},
  \bibinfo{pages}{7341} (\bibinfo{year}{2012}).

\bibitem[{\citenamefont{Tamai et~al.}(2005)\citenamefont{Tamai, Seitsonen,
  Fasel, Shen, Osterwalder, and Greber}}]{Tamai.2005}
\bibinfo{author}{\bibfnamefont{A.}~\bibnamefont{Tamai}},
  \bibinfo{author}{\bibfnamefont{A.~P.} \bibnamefont{Seitsonen}},
  \bibinfo{author}{\bibfnamefont{R.}~\bibnamefont{Fasel}},
  \bibinfo{author}{\bibfnamefont{Z.-X.} \bibnamefont{Shen}},
  \bibinfo{author}{\bibfnamefont{J.}~\bibnamefont{Osterwalder}},
  \bibnamefont{and} \bibinfo{author}{\bibfnamefont{T.}~\bibnamefont{Greber}},
  \bibinfo{journal}{Phys. Rev. B} \textbf{\bibinfo{volume}{72}},
  \bibinfo{pages}{085421} (\bibinfo{year}{2005}).

\bibitem[{\citenamefont{Hoffmann}(1987)}]{Hoffmann.1987}
\bibinfo{author}{\bibfnamefont{R.}~\bibnamefont{Hoffmann}},
  \bibinfo{journal}{{Angew. Chem.}} \textbf{\bibinfo{volume}{26}},
  \bibinfo{pages}{846} (\bibinfo{year}{1987}), ISSN \bibinfo{issn}{0570-0833}.

\bibitem[{\citenamefont{Golze et~al.}(2019)\citenamefont{Golze, Dvorak, and
  Rinke}}]{Golze.2019}
\bibinfo{author}{\bibfnamefont{D.}~\bibnamefont{Golze}},
  \bibinfo{author}{\bibfnamefont{M.}~\bibnamefont{Dvorak}}, \bibnamefont{and}
  \bibinfo{author}{\bibfnamefont{P.}~\bibnamefont{Rinke}},
  \bibinfo{journal}{{Front. Chem.}} \textbf{\bibinfo{volume}{7}},
  \bibinfo{pages}{377} (\bibinfo{year}{2019}), ISSN \bibinfo{issn}{2296-2646}.

\bibitem[{\citenamefont{Perdew et~al.}(1996)\citenamefont{Perdew, Burke, and
  Ernzerhof}}]{Perdew.1996}
\bibinfo{author}{\bibnamefont{Perdew}}, \bibinfo{author}{\bibnamefont{Burke}},
  \bibnamefont{and} \bibinfo{author}{\bibnamefont{Ernzerhof}},
  \bibinfo{journal}{Phys. Rev. Lett.} \textbf{\bibinfo{volume}{77}},
  \bibinfo{pages}{3865} (\bibinfo{year}{1996}).

\bibitem[{\citenamefont{Kresse and Hafner}(1993)}]{Kresse1993}
\bibinfo{author}{\bibfnamefont{G.}~\bibnamefont{Kresse}} \bibnamefont{and}
  \bibinfo{author}{\bibfnamefont{J.}~\bibnamefont{Hafner}},
  \bibinfo{journal}{Phys. Rev. B} \textbf{\bibinfo{volume}{47}},
  \bibinfo{pages}{558} (\bibinfo{year}{1993}).

\bibitem[{\citenamefont{Kresse and Joubert}(1999)}]{Kresse1999}
\bibinfo{author}{\bibfnamefont{G.}~\bibnamefont{Kresse}} \bibnamefont{and}
  \bibinfo{author}{\bibfnamefont{D.}~\bibnamefont{Joubert}},
  \bibinfo{journal}{Phys. Rev. B} \textbf{\bibinfo{volume}{59}},
  \bibinfo{pages}{1758} (\bibinfo{year}{1999}).

\bibitem[{\citenamefont{Tkatchenko and Scheffler}(2009)}]{Tkatchenko2009}
\bibinfo{author}{\bibfnamefont{A.}~\bibnamefont{Tkatchenko}} \bibnamefont{and}
  \bibinfo{author}{\bibfnamefont{M.}~\bibnamefont{Scheffler}},
  \bibinfo{journal}{Phys. Rev. Lett.} \textbf{\bibinfo{volume}{102}},
  \bibinfo{pages}{073005} (\bibinfo{year}{2009}).

\bibitem[{\citenamefont{Bl\"ochl}(1994)}]{Bloechl1994}
\bibinfo{author}{\bibfnamefont{P.~E.} \bibnamefont{Bl\"ochl}},
  \bibinfo{journal}{Phys. Rev. B} \textbf{\bibinfo{volume}{50}},
  \bibinfo{pages}{17953} (\bibinfo{year}{1994}).

\bibitem[{\citenamefont{L{\"u}ftner et~al.}(2017)\citenamefont{L{\"u}ftner,
  Wei{\ss}, Yang, Hurdax, Feyer, Gottwald, Koller, Soubatch, Puschnig, Ramsey
  et~al.}}]{Luftner.2017}
\bibinfo{author}{\bibfnamefont{D.}~\bibnamefont{L{\"u}ftner}},
  \bibinfo{author}{\bibfnamefont{S.}~\bibnamefont{Wei{\ss}}},
  \bibinfo{author}{\bibfnamefont{X.}~\bibnamefont{Yang}},
  \bibinfo{author}{\bibfnamefont{P.}~\bibnamefont{Hurdax}},
  \bibinfo{author}{\bibfnamefont{V.}~\bibnamefont{Feyer}},
  \bibinfo{author}{\bibfnamefont{A.}~\bibnamefont{Gottwald}},
  \bibinfo{author}{\bibfnamefont{G.}~\bibnamefont{Koller}},
  \bibinfo{author}{\bibfnamefont{S.}~\bibnamefont{Soubatch}},
  \bibinfo{author}{\bibfnamefont{P.}~\bibnamefont{Puschnig}},
  \bibinfo{author}{\bibfnamefont{M.~G.} \bibnamefont{Ramsey}},
  \bibnamefont{et~al.}, \bibinfo{journal}{Phys. Rev. B}
  \textbf{\bibinfo{volume}{96}}, \bibinfo{pages}{125402}
  (\bibinfo{year}{2017}).

\bibitem[{\citenamefont{Hasegawa et~al.}(1998)\citenamefont{Hasegawa, Miyamae,
  Yakushi, Inokuchi, Seki, and Ueno}}]{Hasegawa.1998}
\bibinfo{author}{\bibfnamefont{S.}~\bibnamefont{Hasegawa}},
  \bibinfo{author}{\bibfnamefont{T.}~\bibnamefont{Miyamae}},
  \bibinfo{author}{\bibfnamefont{K.}~\bibnamefont{Yakushi}},
  \bibinfo{author}{\bibfnamefont{H.}~\bibnamefont{Inokuchi}},
  \bibinfo{author}{\bibfnamefont{K.}~\bibnamefont{Seki}}, \bibnamefont{and}
  \bibinfo{author}{\bibfnamefont{N.}~\bibnamefont{Ueno}},
  \bibinfo{journal}{Phys. Rev. B} \textbf{\bibinfo{volume}{58}},
  \bibinfo{pages}{4927} (\bibinfo{year}{1998}).

\bibitem[{\citenamefont{Puschnig and L{\"u}ftner}(2015)}]{Puschnig.2015}
\bibinfo{author}{\bibfnamefont{P.}~\bibnamefont{Puschnig}} \bibnamefont{and}
  \bibinfo{author}{\bibfnamefont{D.}~\bibnamefont{L{\"u}ftner}},
  \bibinfo{journal}{J. Electron. Spectrosc. Rel. Phenom.}
  \textbf{\bibinfo{volume}{200}}, \bibinfo{pages}{193} (\bibinfo{year}{2015}).

\bibitem[{\citenamefont{Zhu et~al.}(2006)\citenamefont{Zhu, Dutton, Quinn,
  Lindstrom, Schultz, and Truhlar}}]{Zhu.2006}
\bibinfo{author}{\bibfnamefont{X.-Y.} \bibnamefont{Zhu}},
  \bibinfo{author}{\bibfnamefont{G.}~\bibnamefont{Dutton}},
  \bibinfo{author}{\bibfnamefont{D.~P.} \bibnamefont{Quinn}},
  \bibinfo{author}{\bibfnamefont{C.~D.} \bibnamefont{Lindstrom}},
  \bibinfo{author}{\bibfnamefont{N.~E.} \bibnamefont{Schultz}},
  \bibnamefont{and} \bibinfo{author}{\bibfnamefont{D.~G.}
  \bibnamefont{Truhlar}}, \bibinfo{journal}{Phys. Rev. B}
  \textbf{\bibinfo{volume}{74}}, \bibinfo{pages}{241401(R)}
  (\bibinfo{year}{2006}).

\bibitem[{\citenamefont{Tamai et~al.}(2006)\citenamefont{Tamai, Seitsonen,
  Greber, and Osterwalder}}]{Tamai.2006}
\bibinfo{author}{\bibfnamefont{A.}~\bibnamefont{Tamai}},
  \bibinfo{author}{\bibfnamefont{A.~P.} \bibnamefont{Seitsonen}},
  \bibinfo{author}{\bibfnamefont{T.}~\bibnamefont{Greber}}, \bibnamefont{and}
  \bibinfo{author}{\bibfnamefont{J.}~\bibnamefont{Osterwalder}},
  \bibinfo{journal}{Phys. Rev. B} \textbf{\bibinfo{volume}{74}},
  \bibinfo{pages}{028301} (\bibinfo{year}{2006}).

\bibitem[{\citenamefont{He et~al.}(2006)\citenamefont{He, Arita, Namatame,
  Taniguchi, Li, and Li}}]{He.2006}
\bibinfo{author}{\bibfnamefont{S.}~\bibnamefont{He}},
  \bibinfo{author}{\bibfnamefont{M.}~\bibnamefont{Arita}},
  \bibinfo{author}{\bibfnamefont{H.}~\bibnamefont{Namatame}},
  \bibinfo{author}{\bibfnamefont{M.}~\bibnamefont{Taniguchi}},
  \bibinfo{author}{\bibfnamefont{H.-N.} \bibnamefont{Li}}, \bibnamefont{and}
  \bibinfo{author}{\bibfnamefont{H.-Y.} \bibnamefont{Li}}, \bibinfo{journal}{J.
  Phys. Condens. Matter} \textbf{\bibinfo{volume}{19}}, \bibinfo{pages}{026202}
  (\bibinfo{year}{2006}).

\bibitem[{\citenamefont{Stadtm{\"u}ller
  et~al.}(2019)\citenamefont{Stadtm{\"u}ller, Emmerich, Jungkenn, Haag,
  Rollinger, Eich, Maniraj, Aeschlimann, Cinchetti, and
  Mathias}}]{Stadtmuller.2019}
\bibinfo{author}{\bibfnamefont{B.}~\bibnamefont{Stadtm{\"u}ller}},
  \bibinfo{author}{\bibfnamefont{S.}~\bibnamefont{Emmerich}},
  \bibinfo{author}{\bibfnamefont{D.}~\bibnamefont{Jungkenn}},
  \bibinfo{author}{\bibfnamefont{N.}~\bibnamefont{Haag}},
  \bibinfo{author}{\bibfnamefont{M.}~\bibnamefont{Rollinger}},
  \bibinfo{author}{\bibfnamefont{S.}~\bibnamefont{Eich}},
  \bibinfo{author}{\bibfnamefont{M.}~\bibnamefont{Maniraj}},
  \bibinfo{author}{\bibfnamefont{M.}~\bibnamefont{Aeschlimann}},
  \bibinfo{author}{\bibfnamefont{M.}~\bibnamefont{Cinchetti}},
  \bibnamefont{and} \bibinfo{author}{\bibfnamefont{S.}~\bibnamefont{Mathias}},
  \bibinfo{journal}{Nat. Commun.} \textbf{\bibinfo{volume}{10}},
  \bibinfo{pages}{1470} (\bibinfo{year}{2019}).

\bibitem[{\citenamefont{Koller et~al.}(2007)\citenamefont{Koller, Berkebile,
  Oehzelt, Puschnig, Ambrosch-Draxl, Netzer, and Ramsey}}]{Koller.2007}
\bibinfo{author}{\bibfnamefont{G.}~\bibnamefont{Koller}},
  \bibinfo{author}{\bibfnamefont{S.}~\bibnamefont{Berkebile}},
  \bibinfo{author}{\bibfnamefont{M.}~\bibnamefont{Oehzelt}},
  \bibinfo{author}{\bibfnamefont{P.}~\bibnamefont{Puschnig}},
  \bibinfo{author}{\bibfnamefont{C.}~\bibnamefont{Ambrosch-Draxl}},
  \bibinfo{author}{\bibfnamefont{F.~P.} \bibnamefont{Netzer}},
  \bibnamefont{and} \bibinfo{author}{\bibfnamefont{M.~G.}
  \bibnamefont{Ramsey}}, \bibinfo{journal}{Science}
  \textbf{\bibinfo{volume}{317}}, \bibinfo{pages}{351} (\bibinfo{year}{2007}).

\end{thebibliography}

\end{document}


\title{Signatures of an Atomic Crystal in the Band Structure of a Molecular Thin Film - Supplemental Material}

\author{Norman Haag}
\affiliation{Department of Physics and Research Center OPTIMAS, University of Kaiserslautern, 67663 Kaiserslautern, Germany}
\author{Daniel L\"uftner}
\affiliation{Institute of Physics, University of Graz, NAWI Graz, Universit\"atsplatz 5, 8010 Graz, Austria}
\author{Florian Haag}
\affiliation{Department of Physics and Research Center OPTIMAS, University of Kaiserslautern, 67663 Kaiserslautern, Germany}
\author{Johannes Seidel}
\affiliation{Department of Physics and Research Center OPTIMAS, University of Kaiserslautern, 67663 Kaiserslautern, Germany}
\author{Leah L. Kelly}
\affiliation{Department of Physics and Research Center OPTIMAS, University of Kaiserslautern, 67663 Kaiserslautern, Germany}
\author{Giovanni Zamborlini}
\affiliation{Experimentelle Physik VI, Technische Universit\"at Dortmund, 44221 Dortmund, Germany}
\affiliation{Peter Gr\"unberg Institut (PGI-6), Forschungszentrum J\"ulich, 52425 J\"ulich, Germany}
\author{Matteo Jugovac}
\affiliation{Peter Gr\"unberg Institut (PGI-6), Forschungszentrum J\"ulich, 52425 J\"ulich, Germany}
\author{Vitaliy Feyer}
\affiliation{Peter Gr\"unberg Institut (PGI-6), Forschungszentrum J\"ulich, 52425 J\"ulich, Germany}
\author{Martin Aeschlimann}
\affiliation{Department of Physics and Research Center OPTIMAS, University of Kaiserslautern, 67663 Kaiserslautern, Germany}
\author{Peter Puschnig}
\affiliation{Institute of Physics, University of Graz, NAWI Graz, Universit\"atsplatz 5, 8010 Graz, Austria}
\author{Mirko Cinchetti}
\affiliation{Experimentelle Physik VI, Technische Universit\"at Dortmund, 44221 Dortmund, Germany}
\author{Benjamin Stadtm\"uller}
\affiliation{Department of Physics and Research Center OPTIMAS, University of Kaiserslautern, 67663 Kaiserslautern, Germany}
\email{bstadtmueller@physik.uni-kl.de}

\maketitle

\section{Sample Preparation Procedure}
The C$_{60}$ films were grown in situ on an Ag(111) single crystal. Prior to the deposition of C$_{60}$, the Ag(111) crystal surface was prepared by repeated cycles of argon ion bombardment and subsequent annealing. The quality and cleanness of the Ag(111) surface was confirmed by the existence of well defined diffraction spots in low energy electron diffraction (LEED) with narrow line profiles as well as by the presence of the Shockley surface state in momentum resolved photoemission spectroscopy. The C$_{60}$ films were subsequently grown by molecular beam epitaxy using a commercial Knudsen cell evaporator (Kentax GmbH) at a sublimation temperature of $633\,$K. The film thickness was controlled by evaporation time and molecular flux and verified after the deposition procedure by core level spectroscopy of the C1s and Ag3d levels. In our study, the film thickness was determined to be $(5.0\pm 0.7)\,$ML.
 
\section{Crystal Structure of the C$_{60}$ thin film}

The crystalline structure of the C$_{60}$ thin film was investigated by LEED. An exemplary LEED pattern of this film is shown in Fig.~1a. The best agreement between our LEED data and theoretical simulations was obtained for a superposition of three different structures. The major part of the LEED pattern can be described by a $(2\sqrt{3}\times 2\sqrt{3})R30^\circ$ superstructure in agreement with previous studies \cite{ Shi.2012, Tamai.2005}. The simulated LEED pattern is superimposed onto the experimental data in Fig.~1b as blue circles. In addition, we find diffraction spots of a $(2\sqrt{3}\times 2\sqrt{3})R30^\circ$ superstructure rotated by $\pm18^\circ$ (see LEED simulation in Fig.~1c) and rotated by $\pm 30^\circ$ (see LEED simulation in Fig.~1d). Note that the intensity of the diffraction spots of the $(2\sqrt{3}\times 2\sqrt{3})R30^\circ$$\pm30^\circ$ structure is very low pointing to a marginal relative contribution of this structure to the C$_{60}$ thin film. In our further analysis, the latter domain can hence be neglected.

\begin{figure}[ht]
	\centering
		\includegraphics[height=5.25cm]{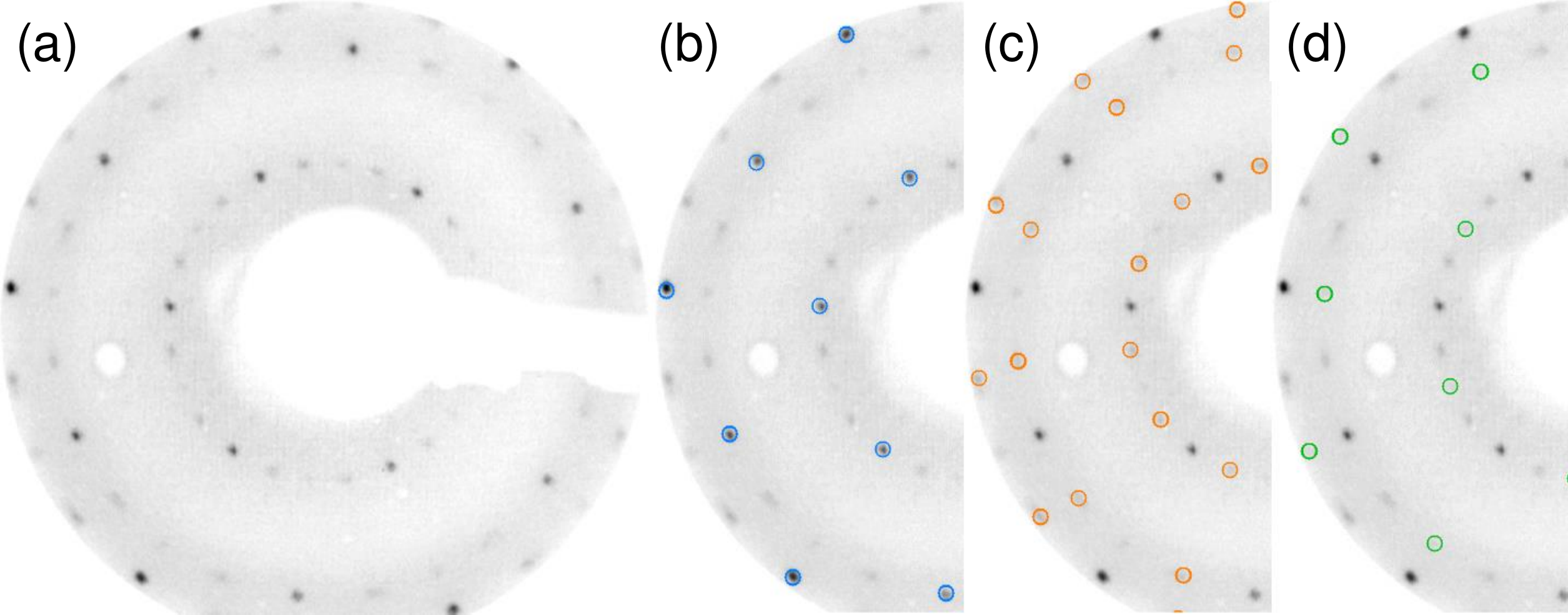}
	\caption{LEED image for a multilayer of C$_{60}$ grown on Ag(111) using a beam energy of $25\,$eV (a).The structure is a superposition of two contributions: a $(2\sqrt{3}\times 2\sqrt{3})R30^\circ$ superstructure (blue circles in (b)) and two domains of a $(2\sqrt{3}\times 2\sqrt{3})R30^\circ$ superstructure rotated by $\pm18^\circ$ (orange circles in (c)). A minor contribution stems from additional domains rotated by $\pm30^\circ$ (green circles in (d)). } 
	\label{fig:Fig1}
\end{figure}

\section{Experimental Methods}
All photoemission experiments were conducted at the NanoEsca end station at the Elettra Synchrotron Light Source, Trieste. The momentum-resolved photoemission yield was recorded with the photoemission electron microscopy system NanoEsca (Focus GmbH) \cite{Kromker.2008} which was operated in k-space mode. All experiments were performed in a fixed experimental geometry, i.e., with a fixed angle of incidence of the synchrotron beam of $65^\circ$ with respect to the surface normal as shown in Fig.~2. Photoemission data were recorded with p- and s-polarized light. For p-polarization the electric field vector (blue arrow) is parallel to the plane of incidence and for s-polarized it is perpendicular to the plane of incidence, i.e., it is located parallel to surface plane.
\begin{figure}[ht]
	\centering
		\includegraphics[width=5cm]{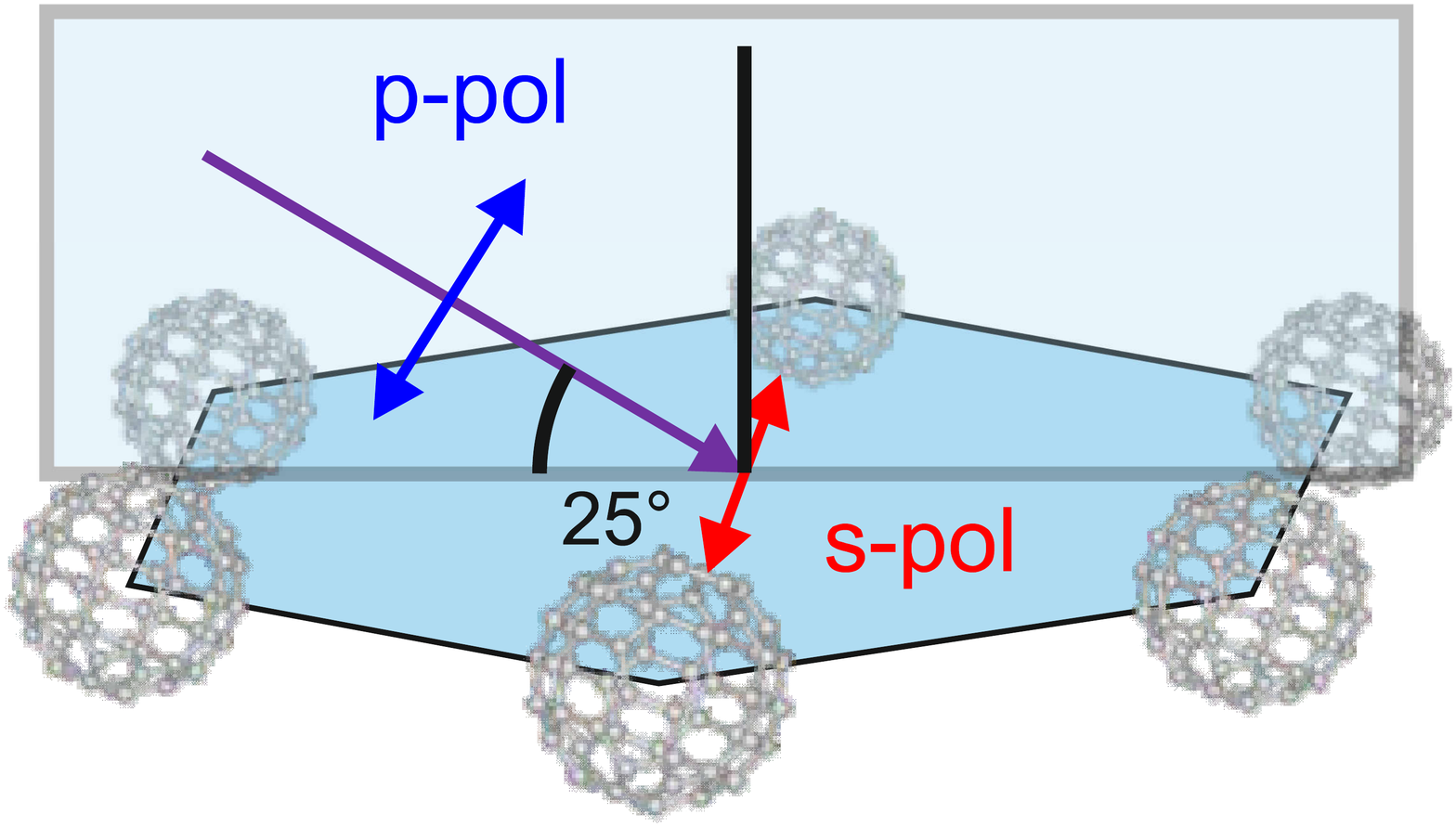}
	\caption{Experimental geometry of the polarization dependent photoemission experiments.} 
	\label{fig:Fig2}
\end{figure}

\section{Computational Methods}
The electronic structure calculations and the simulations of the momentum maps are based on ab-initio computations within the framework of density functional theory (DFT) employing the VASP code \cite{Kresse1993,Kresse1999}. 
The C$_{60}$ film is modeled by a free-standing layer of C$_{60}$-molecules in a hexagonal unit cell with an in-plane lattice parameter of {19.85\,\AA} containing four C$_{60}$ molecules with an additional vacuum layer of about {15\,\AA} in the out-of-plane direction. This structure corresponds to a (111)-cut through the low-temperature bulk crystal structure of C$_{60}$.
For the geometry relaxations of the internal ionic degrees of freedom, the generalized gradient approximation (GGA) 
is used in conjunction with the Perdew-Burke-Ernzerhof (PBE) functional \cite{Perdew1996} and the van-der-Waals corrections according to the Tkatchenko-Scheffler method are added.\cite{Tkatchenko2009}
Using the projector augmented wave (PAW) method,\cite{Bloechl1994} a plane-wave cutoff of $500\,$eV is employed. For $k$-point sampling, a $\Gamma$-centered grid of $8 \times 8 \times 1$ points is used and a first-order Methfessel-
Paxton smearing of $0.1\,$eV is utilized. Based on the relaxed adsorption geometries, we have computed the (projected) density of states. The Kohn-Sham eigenvalues and eigenstates are also the basis for the simulations of the photoemission intensity within the framework of photoemission tomography. Here, we have approximated the final state of the photoemission process by a plane-wave\cite{Luftner.2017} and assumed an inner potential $V_0$ of $13\,$eV\cite{Hasegawa.1998}. For the simulations of the constant binding energy momentum maps and the band maps, an $8 \times 8 \times 4$ sampling of the Brillouin zone and Gaussian broadenings of $0.05\,$\AA$^{-1}$ and $0.1\,$eV in the momentum and energy axes have been chosen, respectively. 

\section{Rotational Domains in Photoemission Tomography Simulations}

\begin{figure}[ht]
	\centering
		\includegraphics[height=9cm]{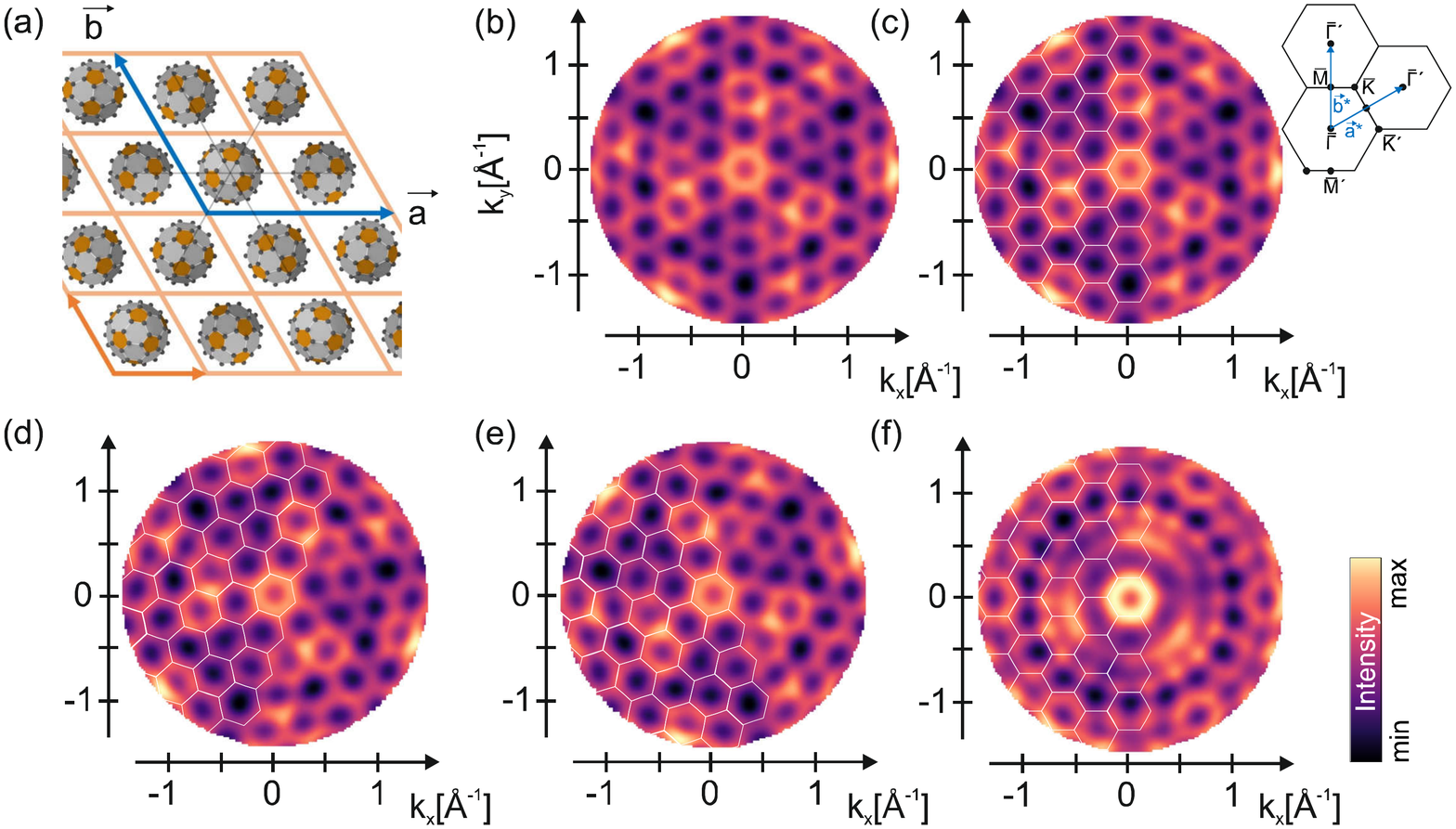}
	\caption{(a) Structural model of the $(2\sqrt{3}\times 2\sqrt{3})R30^\circ$ structure with four C$_{60}$ molecules per unit cell. (b) Constant energy map of the $(2\sqrt{3}\times 2\sqrt{3})R30^\circ$ structure at E$_{\mathrm{B}}=3.5\,$eV simulated by PT. The same CE map is superimposed with the lattice of the surface Brillouin zones of the C$_{60}$ structure in (c). The CE maps of the $(2\sqrt{3}\times 2\sqrt{3})R30^\circ$$\pm18^\circ$ are shown in (d) and (e). (f) Total momentum-resolved photoemission yield calculated by adding up the CE maps of the different structural domains.} 
	\label{fig:Fig2}
\end{figure}

The high accuracy of our photoemission tomography (PT) simulations for the valence bands of C$_{60}$ relies on a proper treatment of the additional structural domains observed in our LEED data. The band structure calculation and the PT simulations were performed for a freestanding  $(2\sqrt{3}\times 2\sqrt{3})R30^\circ$ structure with four C$_{60}$ molecules per unit cell, see section \textit{Computational Methods} above. A structural model of the unit cell used in the simulations is shown in Fig.~3a.  An exemplary constant energy (CE) map of the simulated momentum resolved photoemission yield is shown in Fig.~3b for one energy within the HOMO-1 band (E$_{\mathrm{B}}=3.5\,$eV). This binding energy corresponds to one of the binding energy of the HOMO-1 CE maps discussed in Fig.~2 of the main manuscript. The CE map consists of a regular arrangement of hexagonal emission features which follow the periodicity of the $(2\sqrt{3}\times 2\sqrt{3})R30^\circ$ superstructure in momentum space. The different hexagonal emission pattern represent the C$_{60}$ valence band structure in higher surface Brillouin zones as clearly visible in Fig.~3c where the surface Brillouin zones are superimposed onto the same CE map as white hexagons. The directions and high symmetry points of the surface Brillouin zones are indicated in the inset.\\

\begin{figure}[ht]
	\centering
		\includegraphics[height=9cm]{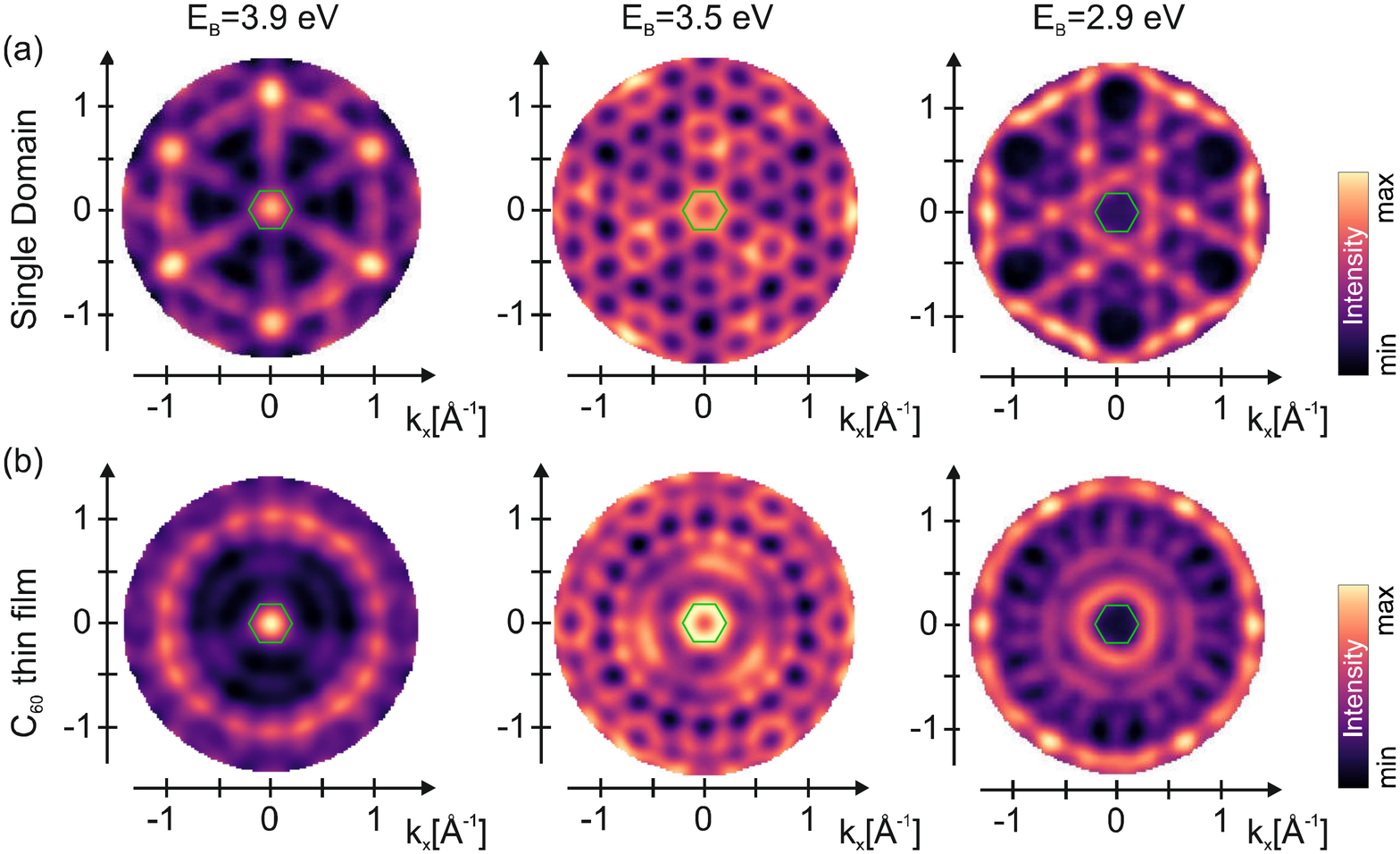}
	\caption{(a) The simulated momentum resolved photoemission yield of the $(2\sqrt{3}\times 2\sqrt{3})R30^\circ$ at E$_{\mathrm{B}}=3.9\,$eV, $3.5\,$eV and $2.9\,$eV. (b) Total momentum-resolved photoemission yield of the C$_{60}$ thin film at the same binding energies.} 
	\label{fig:Fig4}
\end{figure}

The momentum resolved photoemission yield of the $(2\sqrt{3}\times 2\sqrt{3})R30^\circ$$\pm18^\circ$ superstructure can be obtained by rotating the CE maps of the $(2\sqrt{3}\times 2\sqrt{3})R30^\circ$ (Fig.~3b) superstructure by $\pm18^\circ$, see Fig.~3d and e. The total momentum resolved photoemission yield is finally simulated by adding up the contributions of the three structural domains: 
\begin{equation} I(k_x,k_y,E_B)=\alpha \times I_{ 0^\circ} (k_x,k_y,E_B) + \beta\times \left (I_{ +18^\circ} (k_x,k_y,E_B)+ I_{-18^\circ} (k_x,k_y,E_B) \right ) \end{equation}
Here, $\alpha$, and $\beta$ denote the relative contributions of the two different structural domains. The best agreement between our PT simulations and the experiment was obtained for an almost equal ratio of the $\pm0^\circ$ and the $\pm18^\circ$ domains with $\alpha=1$ and $\beta=0.9$. The corresponding CE map is shown in Fig.~3f and in Fig.~2 of the main manuscript. Note that no spectroscopic signature of the $(2\sqrt{3}\times 2\sqrt{3})R30^\circ$$\pm30^\circ$ structure was observed in our momentum-resolved photoemission data. This is in line with the extremely weak intensity of the diffractions spots of this particular rotational domain in our LEED data discussed above. We therefore neglect any contribution of the $(2\sqrt{3}\times 2\sqrt{3})R30^\circ$$\pm30^\circ$ superstructure in our PT simulations. 

The same procedure is repeated for the second binding energy of the valence band structure shown in Fig.~2 of the main manuscript. The momentum-resolved photoemission yield of the $(2\sqrt{3}\times 2\sqrt{3})R30^\circ$ superstructure as well as the total photoemission yield including both rotated domains is shown in Fig.~4 for three characteristic binding energies of the HOMO-1 band.

\newpage
\section{Challenges in Photoemission Tomography Simulations of Localized Molecular Orbitals}

\begin{figure}[ht]
	\centering
		\includegraphics[height=6cm]{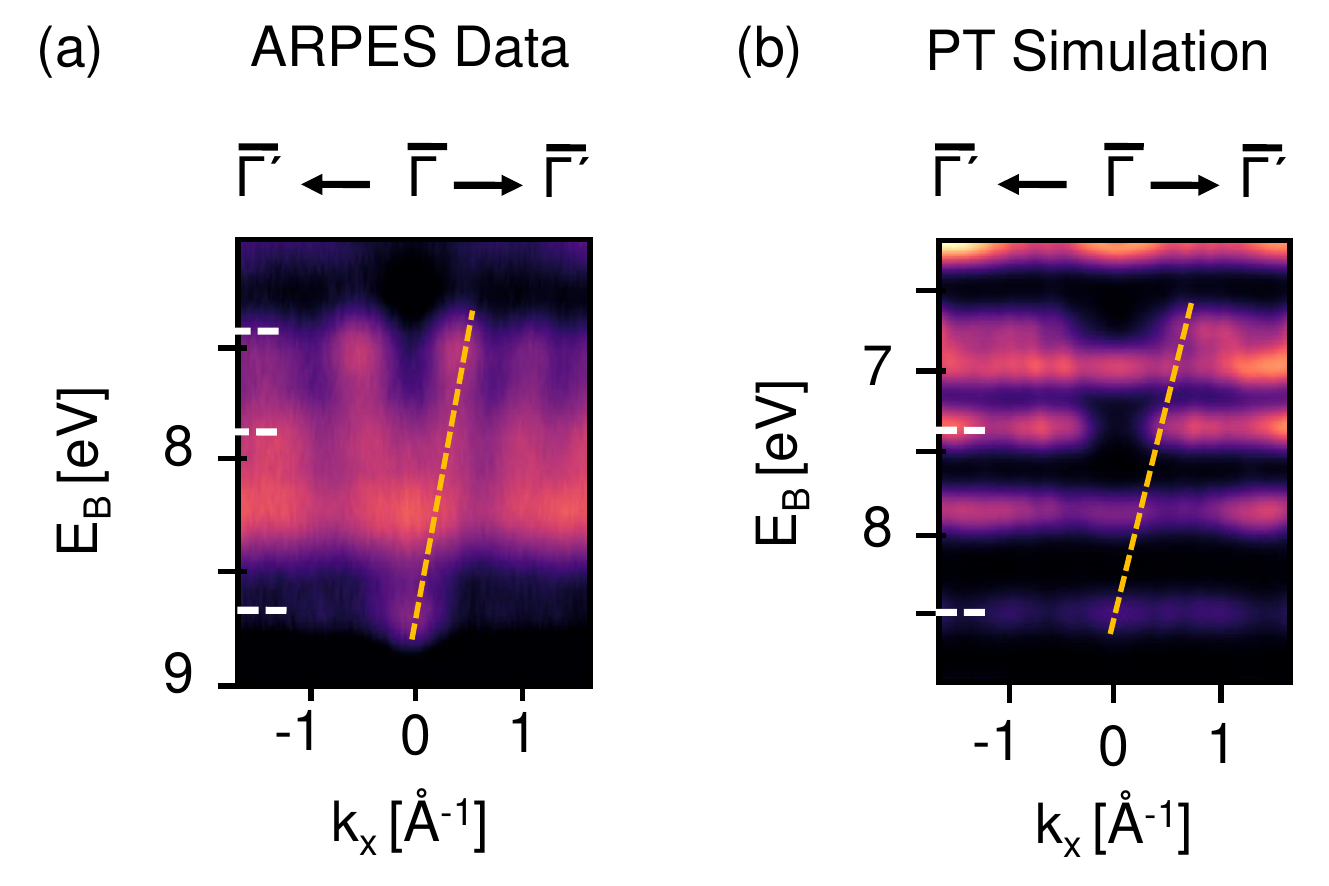}
		\caption{Experimental (a) and simulated (b) energy vs. momentum cut through the 3D ARPES data cube extracted along the $\overline{\Gamma}$~$\overline{\mathrm{M}}$~$\overline{\Gamma}'$-direction of the surface Brillouin zone of a crystalline C$_{60}$ thin film in the binding energy range of the $\sigma$-states. The yellow line indicates the intramolecular band dispersion as guide-to-the-eye.} 
	\label{fig:Fig5}
\end{figure}

Despite the overall excellent qualitative agreement between our momentum-resolved photoemission data and the PT simulations of the localized $\sigma$-state of C$_{60}$ at large binding energies E$_{\mathrm{B}}>5\,$eV, there are also minor but distinct deviations between experiment and simulation in Fig.~4 of the main manuscript. For instance, we observe a slightly different energy and momentum position of the molecular $\sigma$- states in the energy vs. momentum cuts in Fig.~5 leading to a different slope of the almost linear intramolecular dispersion curve. This discrepancy also coincides with a different radius and relative photoemission intensity of the concentric molecular emission features for different E$_\mathrm{B}$ in experiment and theory. We attribute these deviations to the strong k$_\mathrm{\bot}$ dependency of the 3D Fourier transform of the localized molecular states of non-planar molecules. 
To support this conclusion, we depict the k$_\mathrm{\bot}$ dependency of the 3D Fourier transform of the highest occupied molecular orbital (HOMO) of a free C$_{60}$ molecule in Fig.~6. Fig.~6a shows a 2D cut through the 3D Fourier transform of the C$_{60}$ HOMO in the k$_\mathrm{||}$- k$_\mathrm{\bot}$ plane. This 2D cut already illustrates the complex intensity pattern of the 3D Fourier transform with which varies as a function of both k$_\mathrm{||}$ and k$_\mathrm{\bot}$. Consequently, small variations of the total momentum k$_\mathrm{final}$ of the electrons in the photoemission final state result in a different spherical cut through the 3D Fourier transform of the localized molecular orbitals. This in turn severely alters the theoretically predicted CE emission characteristics of the HOMO as shown for three exemplary CE maps of the HOMO simulated for three different momentum k$_\mathrm{final}$ in Fig.~6b. 
In contrast, the k$_\mathrm{\bot}$ dependency of the 3D Fourier transform is almost neglectable for planar molecules.  Fig.~6c shows a k$_\mathrm{||}$- k$_\mathrm{\bot}$ cut through the 3D Fourier transform of the HOMO of the planar molecule PTCDA. The 3D Fourier transformed reveals only a weak intensity modulation along the k$_\mathrm{\bot}$-direction. Consequently, the simulated momentum-dependent photoemission yield of the PTCDA HOMO is almost independent of the total electron momentum k$_\mathrm{final}$ in the photoemission final state k$_\mathrm{final}$. The latter was recently demonstrated experimentally by Weiss et al. for PTCDA/Ag(110)\cite{Weiss.2015}.
This comparison clearly underlines the crucial role of the final state momentum k$_\mathrm{final}$ for the PT of 3D molecules. The latter can be influenced either by the experimental uncertainty of the inner potential V$_0$ of the material or by small deviations of the initial state energy of molecular states in the band structure calculations and the experiment. In the case of C$_{60}$, the self-interaction errors in the band structure calculations result in a significant shift of the $\sigma$-states with respect to the experiment. We hence propose that this effect is responsible for the qualitative difference observed for the PT simulations and the experimentally obtained CE maps of the C$_{60}$ $\sigma$-states. 

\begin{figure}[ht]
	\centering
		\includegraphics[height=9cm]{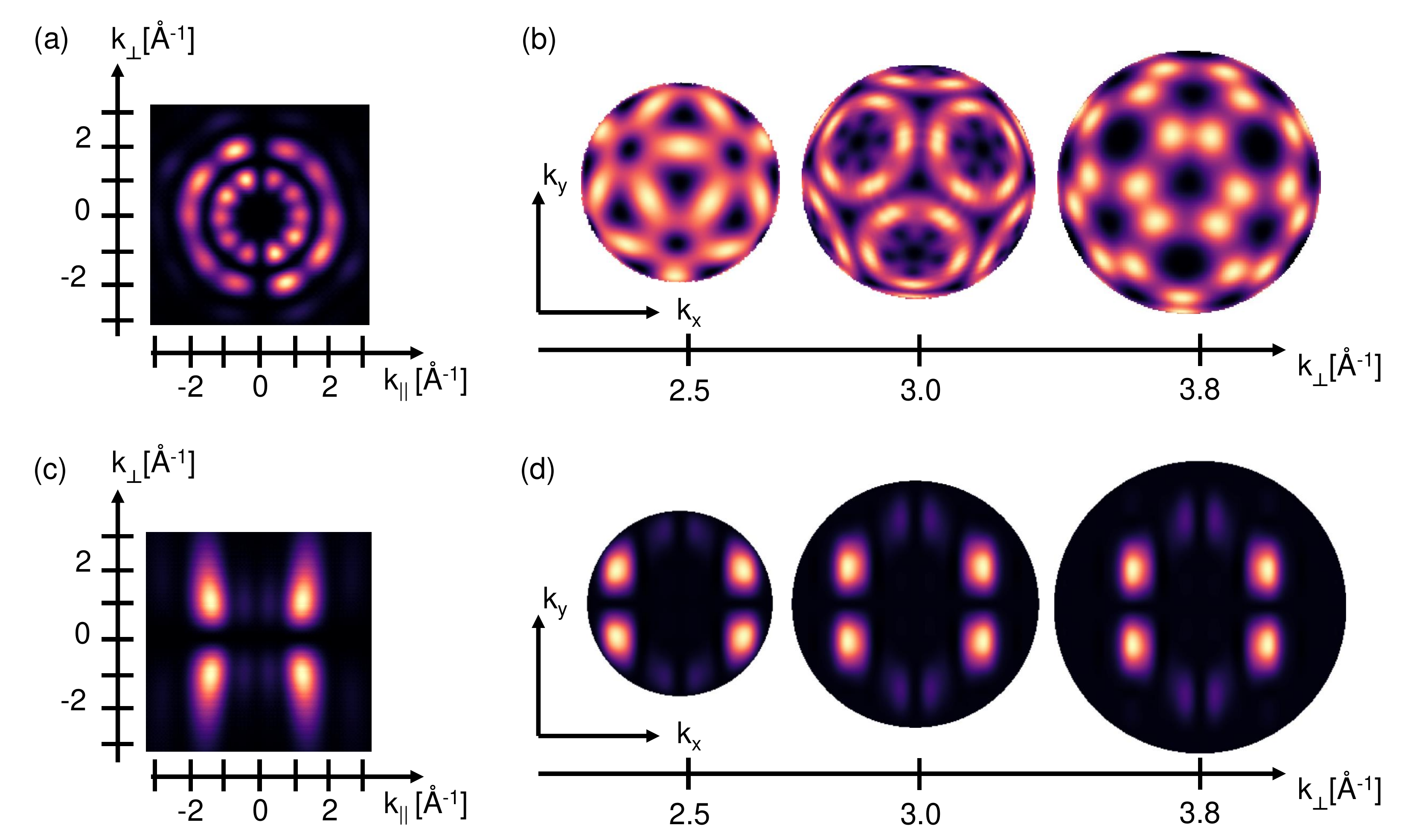}
		\caption{(a) 2D k$_\mathrm{||}$ - k$_\mathrm{\bot}$ cut through the 3D Fourier transform of the C$_{60}$ HOMO. (b) Momentum-dependent photoemission yield of the C$_{60}$ HOMO for different total momentum k$_\mathrm{final}$ of the electrons in the photoemission final state. (c) and (d) show similar plots for the HOMO of the planar model molecule PTCDA.} 
	\label{fig:Fig6}
\end{figure}